\documentclass{JHEP3}
\usepackage{bm}
\usepackage{amsmath}
\usepackage{epsfig}

\title{Symmetry of the CMB sky as a new test of its statistical isotropy. Non Cosmological Octupole?}
\author{P. Naselsky, M. Hansen, J. Kim\\
Niels Bohr Institute and DISCOVERY center, Blegdamsvej 17, 2100 Copenhagen, {\O}, Denmark}

\abstract{
In this article we propose a novel test for statistical anisotropy of the CMB $\Delta T\left(\hat{\mathbf n}=(\theta,\phi)\right)$. The test is based on the fact, that the Galactic foregrounds have a remarcably strong symmetry with respect to their antipodal points $S1:=\hat{\mathbf n}\leftrightarrow -\hat{\mathbf n},\hat{\mathbf n}=(\theta,\phi) $ and $S2:=\hat{\mathbf n}\leftrightarrow \overline{\hat{\mathbf n}},\overline{\hat{\mathbf n}}=\pi-\theta,\phi$ with respect to the Galactic plane, while the cosmological signal should not be symmetric or asymmetric under these transitions. \\
We have applied the test for the octupole component of the WMAP ILC 7 map, by looking at $a_{3,1}$ and $a_{3,3}$, and their ratio to $a_{3,2}$ both for real and imaginary values. We find abnormal symmetry of the octupole component at the level of 0.58\%, compared to Monte Carlo simulations. By using the analysis of the phases of the octupole we found remarkably strong cross-correlations between the phases of the kinematic dipole and the ILC 7 octupole, in full agreement with previous results. \\
We further test the multipole range $2<l<100$, by investigating the ratio between the $l+m=even$ and $l+m=odd$ parts of power spectra. We compare the results to simulations of a Gaussian random sky, and find significant departure from the statistically isotropic and homogeneous case, for a very broad range of multipoles. We found that for the most prominent peaks of our estimator, the phases of the corresponding harmonics are coherent with phases of the octupole. We believe, our test would be very useful for detections of various types of residuals of the foreground and systematic effects at a very broad range of multipoles $2\le l\le 1500-3000$ for the forthcoming PLANCK CMB map, before any conclusions about primordial non-Gaussianity and statistical anisotropy of the CMB.
}

\makeindex
\preprint{-}
\begin{document}

\newcommand{\beq}{\begin{equation}}
\newcommand{\eeq}{\end{equation}}
\newcommand{\be}{\begin{eqnarray}}
\newcommand{\ee}{\end{eqnarray}}
\newcommand{\num}{\nu_\mu}
\newcommand{\nue}{\nu_e}
\newcommand{\nut}{\nu_\tau}
\newcommand{\nus}{\nu_s}
\newcommand{\mnus}{M_s}
\newcommand{\taus}{\tau_{\nu_s}}
\newcommand{\nnt}{n_{\nu_\tau}}
\newcommand{\rnt}{\rho_{\nu_\tau}}
\newcommand{\mnt}{m_{\nu_\tau}}
\newcommand{\tnt}{\tau_{\nu_\tau}}
\newcommand{\rar}{\rightarrow}
\newcommand{\lar}{\leftarrow}
\newcommand{\lrar}{\leftrightarrow}
\newcommand{\dm}{\delta m^2}
\newcommand{\mpl}{m_{Pl}}
\newcommand{\mbh}{M_{BH}}
\newcommand{\nbh}{n_{BH}}
\newcommand{\crit}{{\rm crit}}
\newcommand{\ini}{{\rm in}}
\newcommand{\cmb}{{\rm cmb}}
\newcommand{\rec}{{\rm rec}}

\newcommand{\Odm}{\Omega_{\rm dm}}
\newcommand{\Ob}{\Omega_{\rm b}}
\newcommand{\Om}{\Omega_{\rm m}}
\newcommand{\nb}{n_{\rm b}}
\def\simlt{\lesssim}
\def\simgt{\gtrsim}
\def\Cl{C_{\ell}}
\def\out{{\rm out}}
\def\in{{\rm in}}
\def\mean{{\rm mean}}
\def\zrec{z_{\rm rec}}
\def\zreio{z_{\rm reion}}
\def\wmap{{\it WMAP}} 
\def\planck{{\it Planck}}

%\maketitle

\section{Introduction}
Since the release of the COBE temperature map of the Cosmic Microwave Background (CMB), and after the Wilkinson Microwave Anisotropy Probe (WMAP) (\cite{WMAP1}, \cite{WMAP7}), the modern cosmology has been dramatically changed. Two fundamental hypotheses, the existence of the Dark Matter and the Dark Energy in nature, have found experimental confirmation in the CMB data. Moreover, the power spectrum of the temperature anisotropy and polarization of the CMB is in excellent agreement with theoretical predictions of the simplest models of inflation, providing a remarkably important connection between the physics of the modern Universe and the very beginning of the cosmological creation of the matter.\\
However, right after the COBE mission, some anomalies of the CMB sky attracted very serious attention, starting with the low amplitude of the quadrupole \cite{Oliveira-Costa1}, the quadrupole-octupole alignment with planarity and lack of correlations at $\theta>60^o$ \cite{Copi1}, \cite{Copi3}, \cite{Copi5}, the phase correlations between different multipoles \cite{Chiang}, \cite{naselsky}, \cite{Gorski1}, \cite{Gorski2}, \cite{Gorski3}, the existence of the cold spot \cite{Coldspot1},\cite{Cruz2}, \cite{Cruz3}, the dominance of the power of odd multipoles over even one \cite{Land}, \cite{Jkim1}, \cite{odd_bolpol}, \cite{Universe_odd}, \cite{Burigana},\cite{Jkim3} and \cite{Mhansen}, the ecliptic and Galactic north-south asymmetry of the power spectrum,\cite{Eriksen}, etc (see for review \cite{WMAP7a}. 

In this article, we present a new test for statistical asymmetry and possible non-gaussianity of the CMB, based on an investigation of the symmetries of the CMB sky with respect to different directions on the sphere. This symmetry test will reflect the properties of the Galactic foregrounds and possible systematic effects. The idea of the test is based on the assumption that the primordial CMB signal, unlike any other non-cosmological signal (like the galactic synchrotron, free-free, dust emission and galactic and extragalactic point-like sources), should have no particular symmetries or asymmetries of angular distribution on the sky. This is what one would expect for a completely random, chaotic distribution. At the same time, for instance, the galactic foregrounds clearly have a well defined symmetry of the temperature distribution with respect to their antipodes, and slightly broken, but still ``visible'' symmetry with respect to the Galactic plane \cite{naselsky1}. The instrumental noise, the residuals of the calibrations and possible other effects of systematics have a preferable symmetry in the direction of the north and south ecliptic poles (in Galactic coordinates).

All these symmetries manifest themselves in the coefficients $a_{l,m}$ of the spherical harmonic decomposition of the signal $\Delta T(\theta,\phi)$ trough the even and odd multipoles (the symmetry $S1=\hat{\mathbf n}\leftrightarrow -\hat{\mathbf n},\hat{\mathbf n}=(\theta,\phi) $ with respect to the antipodes), trough the symmetry $S2=\hat{\mathbf n}\leftrightarrow \overline{\hat{\mathbf n}}$ with respect to the Galactic plane, where $\overline{\hat{\mathbf n}}=(\pi-\theta, \phi)$, and trough the symmetry $S3=\hat{\mathbf n}\leftrightarrow \hat{\mathbf m}$, where $\hat{\mathbf m}=(\theta,2\pi-\phi)$.\\
Note, that the octupole component is the most powerful tail of the CMB power spectrum. Before our analysis, the peculiarity of the octupole was widely discussed in connection with quadrupole- octupole alignment \cite{Oliveira-Costa1},\cite{Copi1},\cite{Copi5}. In this article, we take a closer look at the various ratios between the $a_{l,m}$-values making up the octupole, unconnected with any investigation of the quadropole, and find a significant deviation from a random distribution. Further, we create power spectra from the real and imaginary part of the $a_{l,m}$-parameter, where the sum is over even and odd values of $l+m$ only ($D^{+}(l)$ and $D^{-}(l)$ respectively). Then we test the ratio between real and imaginary $D^{\pm}(l)$-values, for a large range of multipoles. The new element of our analysis, which was never discussed before, is that the symmetry of $l=7$ mode of the ILC 7 map is peculiar at the level of 3 events from $10^3$ realizations for $l+m=odd$. For $l+m=even$ the most impressive result is connected with $l=14$, also with 3 events from $10^3$ realizations, while for $l=38$ we have the corresponding probability about $0.6\%$.

It would be worth to note, that our analysis is based on the ILC 7 map, which is contaminated by the point-like sources (Galactic and extra-galactic, residuals of the diffuse foregrounds, uncertainties of the antenna beam and 
possible effects of systematics). This is why the ILC map can not be used for evaluation of the CMB power spectrum
without implementation of the mask (for instance, the KQ75 WMAP mask). After that, the analysis of the $a_{l,m}$-coefficients can not be done, due to a very strong coupling between the coefficients, induced by the mask. However, our method is especially useful for estimating the degree of contamination of the ILC map - in combination with standard methods for determining the power spectrum $C(l)$ from the masked sky - in order to check out possible sources of peculiarities of $C(l)$. We believe, that this method would be especially useful for the ongoing PLANCK mission, where the ILC map would be applicable to a very broad range of multipoles, compared to the WMAP range. An important point is, that for very high $l$, actually, we do not have any theoretical predictions about potentially dangerous zones, which need to be masked. 

There is one more important implementation of our method, related to the non-Gaussianity test (the so called $f_{nl}$- approach), allowing one to constrain different models of inflation. Quadratic corrections to the linear theory of perturbations from inflation (local $f_{nl}$ models), are characterized by a coupling between low and high multipoles, which provides fingerprints of this particular type of non-Gaussianity in the bi-spectrum, and in higher order moments (see for review \cite{komatsu}). To assess the $f_{nl}$ approach with maximal precision, it is clear that we have to detect all possible sources of non-Gaussian contamination of the CMB map, amongst which the point sources seems to be a major component.

The outline of the paper is the following. In section 2 we introduce the various symmetries of the CMB sky, and analyze the ratio between various values of $a_{l,m}$ for the octupole (Section 3). In addition, in Section 3 we will show that the symmetry test is closely connected with the phases of the coefficients of the spherical harmonic decomposition, indicating the most peculiar components. Further,in Section 4 we introduce the pathfinder of peculiar multipoles, based on the symmetry test, created from even and odd values of $l+m$, and finally we apply that method to the ILC 7 map. In section 5, we summarize the results of simulations and comparison with the data, given by our estimators.

\section{The symmetry test}
The temperature fluctuations on the CMB sky can be decomposed into spherical harmonics in the following standard way: 
\begin{eqnarray}
\Delta T(\hat{\mathbf n}) &=& \sum_{l=0}^{l_{max}} \sum_{m=-l}^{l} a_{l,m} Y_{l,m}(\hat{\mathbf n}) = \frac{1}{\sqrt{2\pi}}\sum_{l=2}^{l_{max}}\sqrt{\frac{2l+1}{2}}\Re( a_{l,m=0})P_l(\cos\theta)+\nonumber\\
&+&\frac{2}{\sqrt\pi}\sum_{l=2}^{l_{max}}\sum_{m=1}^{l}\sqrt{\frac{(2l+1)(l-m)!}{2(l+m)!}}P^m_l(\cos\theta) \times \left[\Re(a_{l,m})\cos(m\phi)-\Im(a_{l,m})\sin(m\phi)\right]\nonumber\\
\label{eq1}
\end{eqnarray}
where $a_{l,m}$ is the coefficient of decomposition, and $\hat{\mathbf n} = (\theta, \phi)$, with $\theta$ and $\phi$ being the polar and azimuthal angles on the sky respectively. $\Re$ and $\Im$ denote the real and imaginary parts of the $a_{l,m}$-coefficients, $P^m_l(\cos\theta)$ are the associated Legendre polynomials, and $P_l(\cos\theta)=P^0_l(\cos\theta)$.
\FIGURE{
 \centerline{\includegraphics[scale=.185]{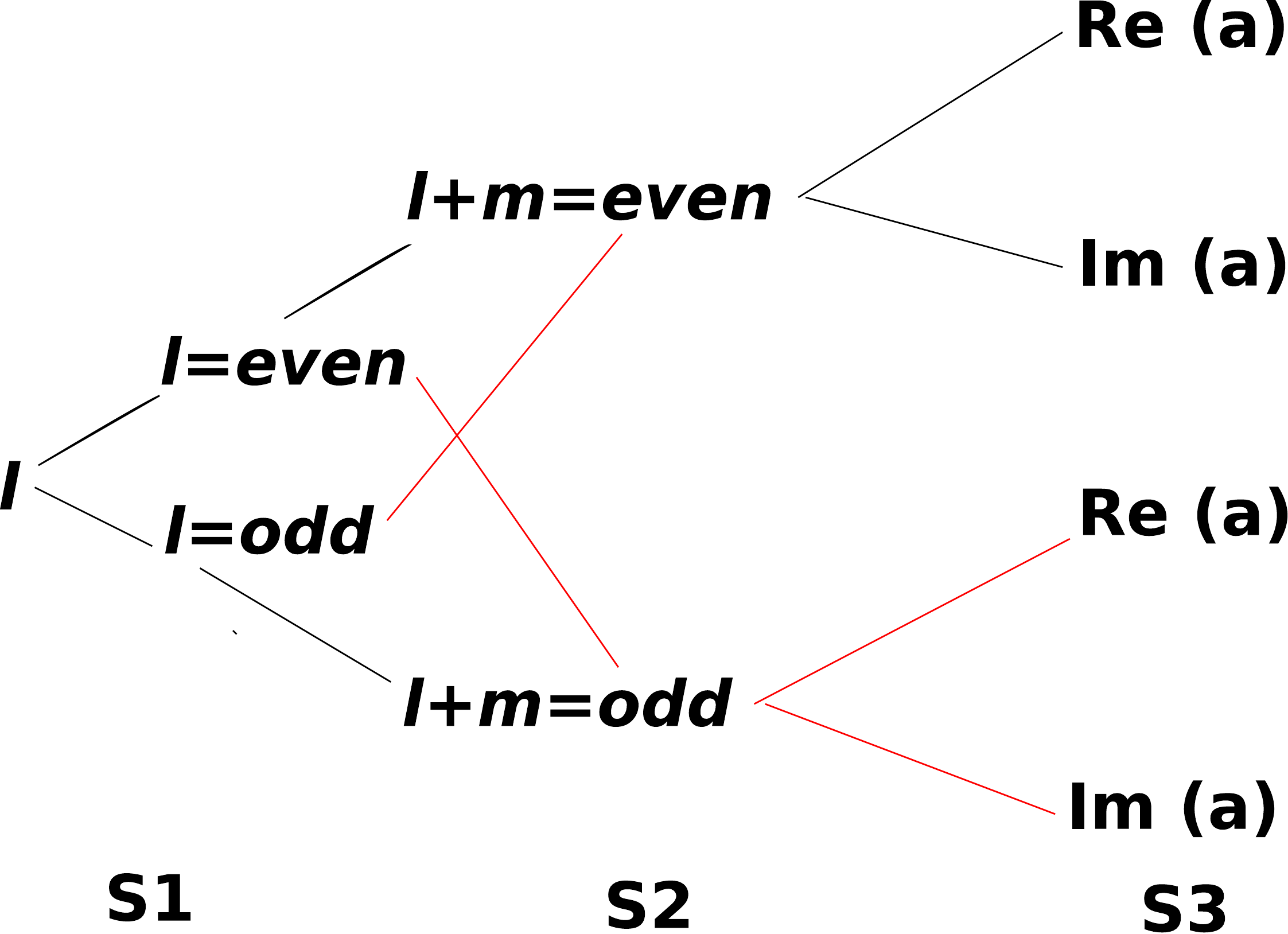}
\includegraphics[scale=.2]{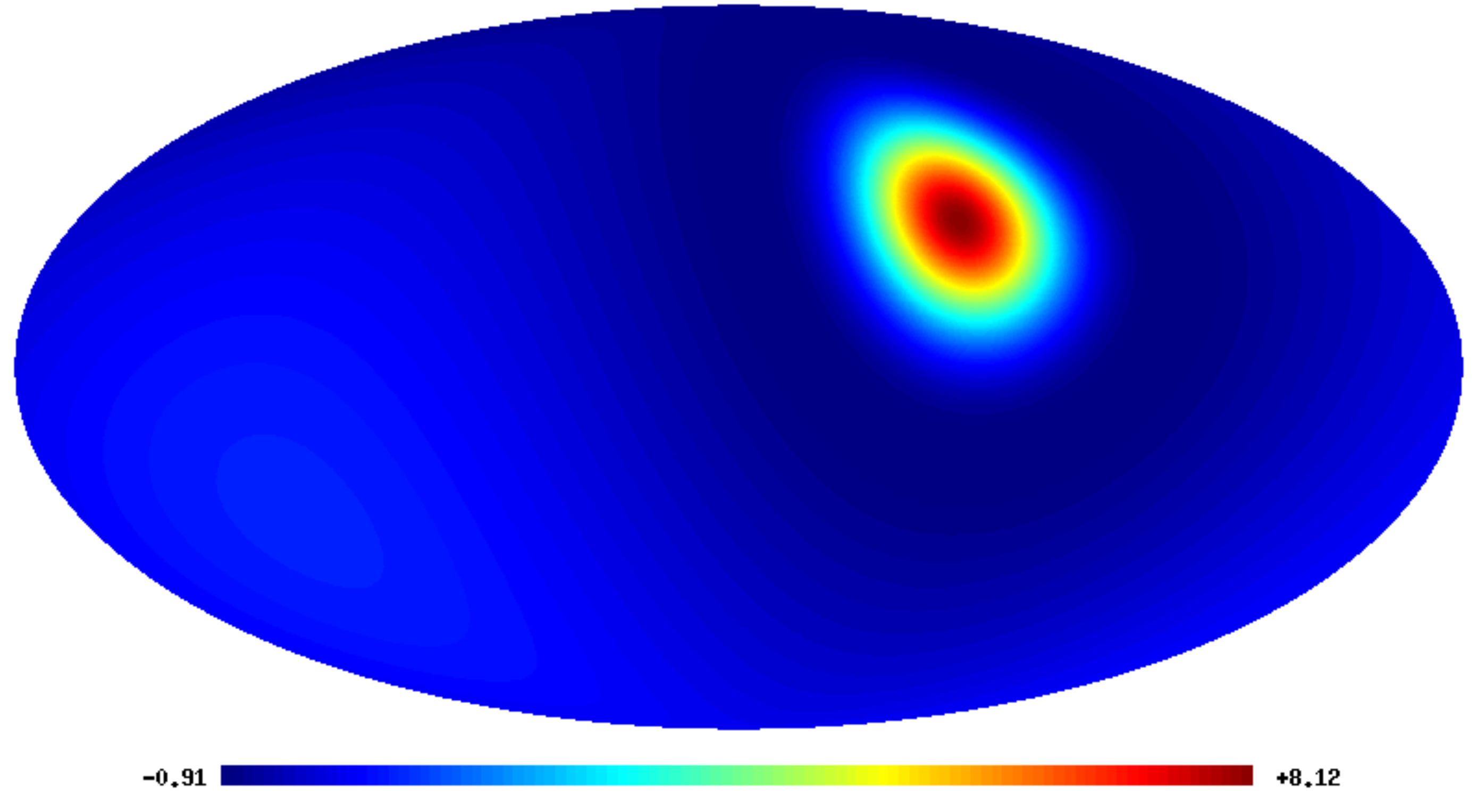}}
\centerline{\includegraphics[scale=.2]{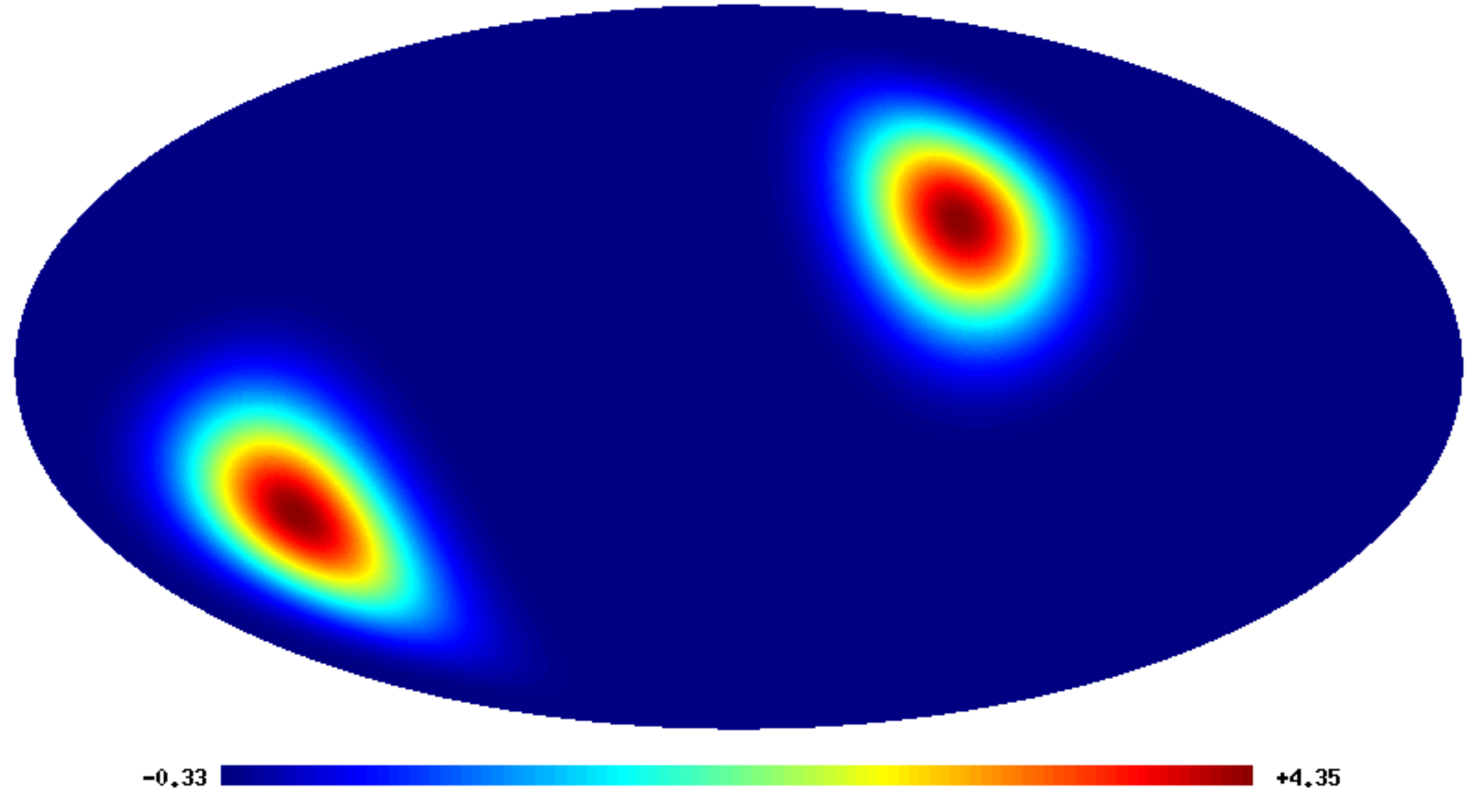}
  \includegraphics[scale=.2]{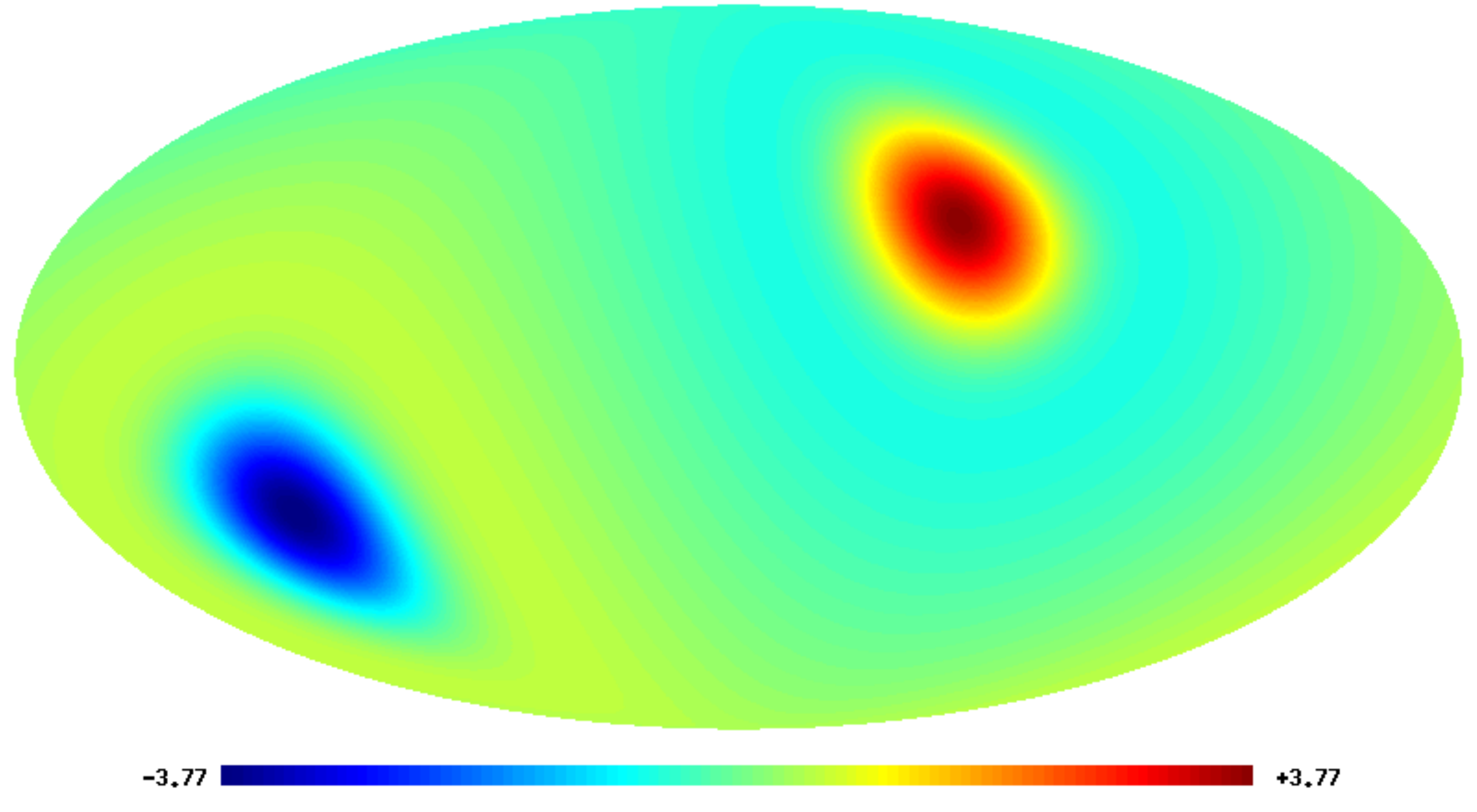}}
\centerline{\includegraphics[scale=.2]{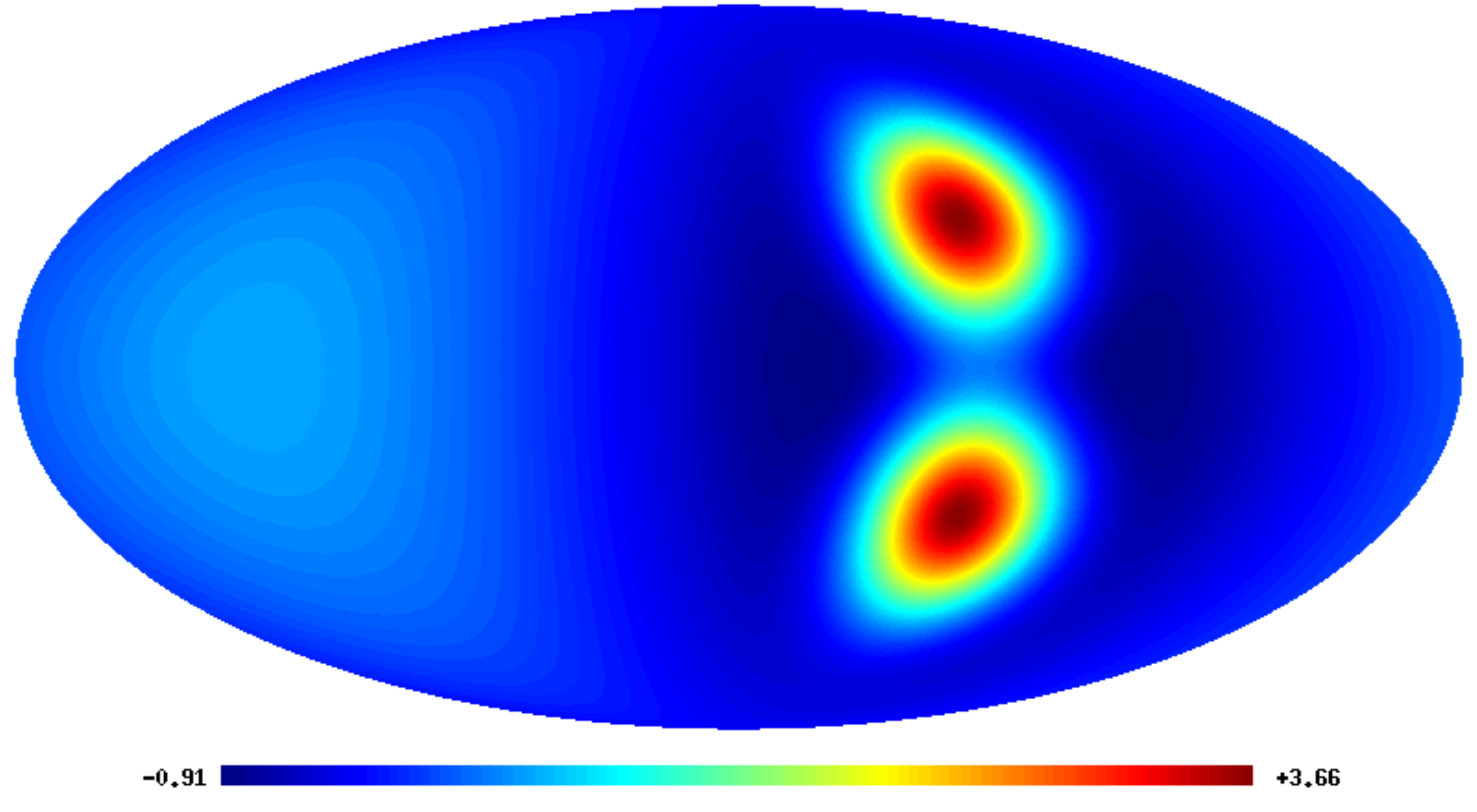}
\includegraphics[scale=.2]{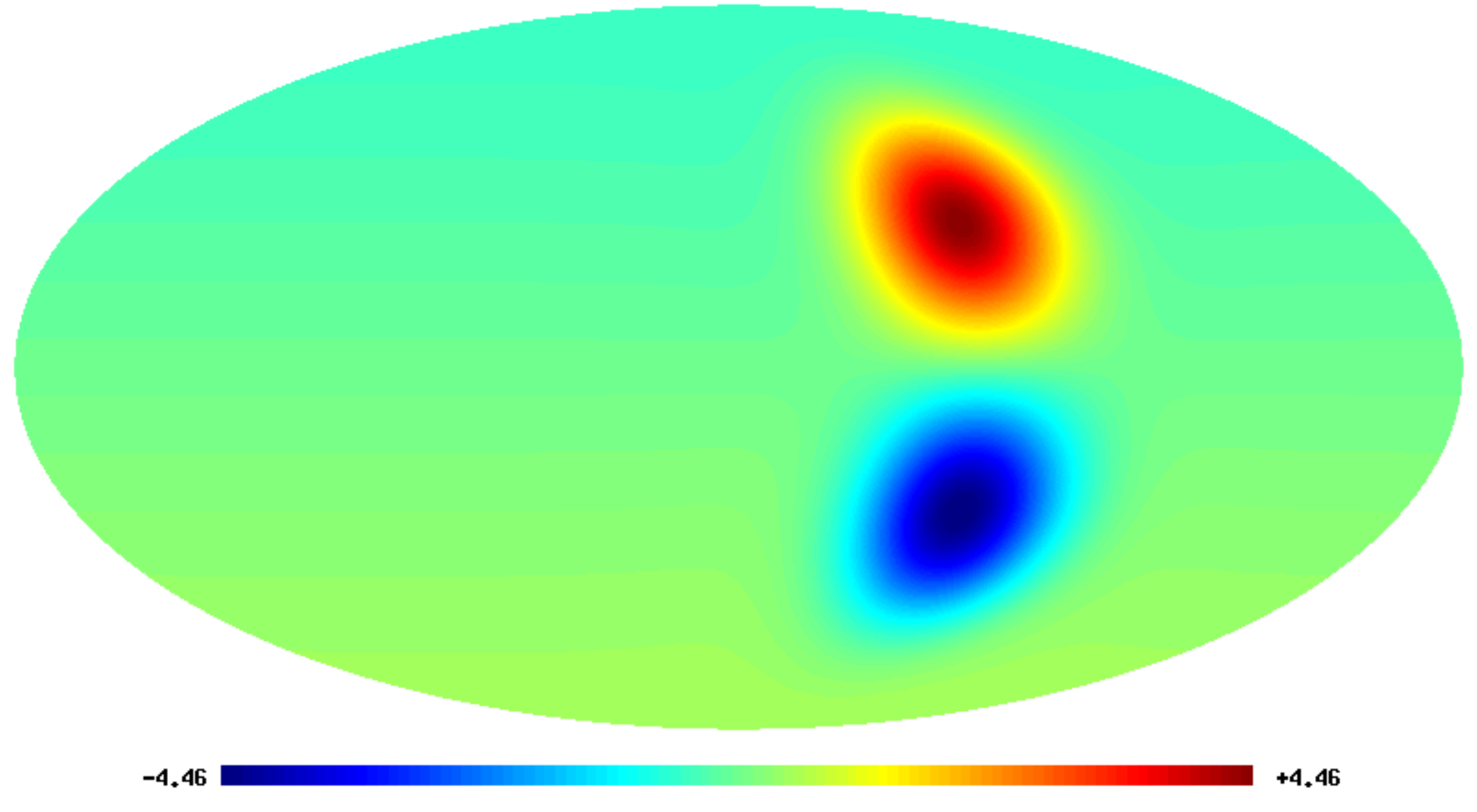}}
\centerline{\includegraphics[scale=.2]{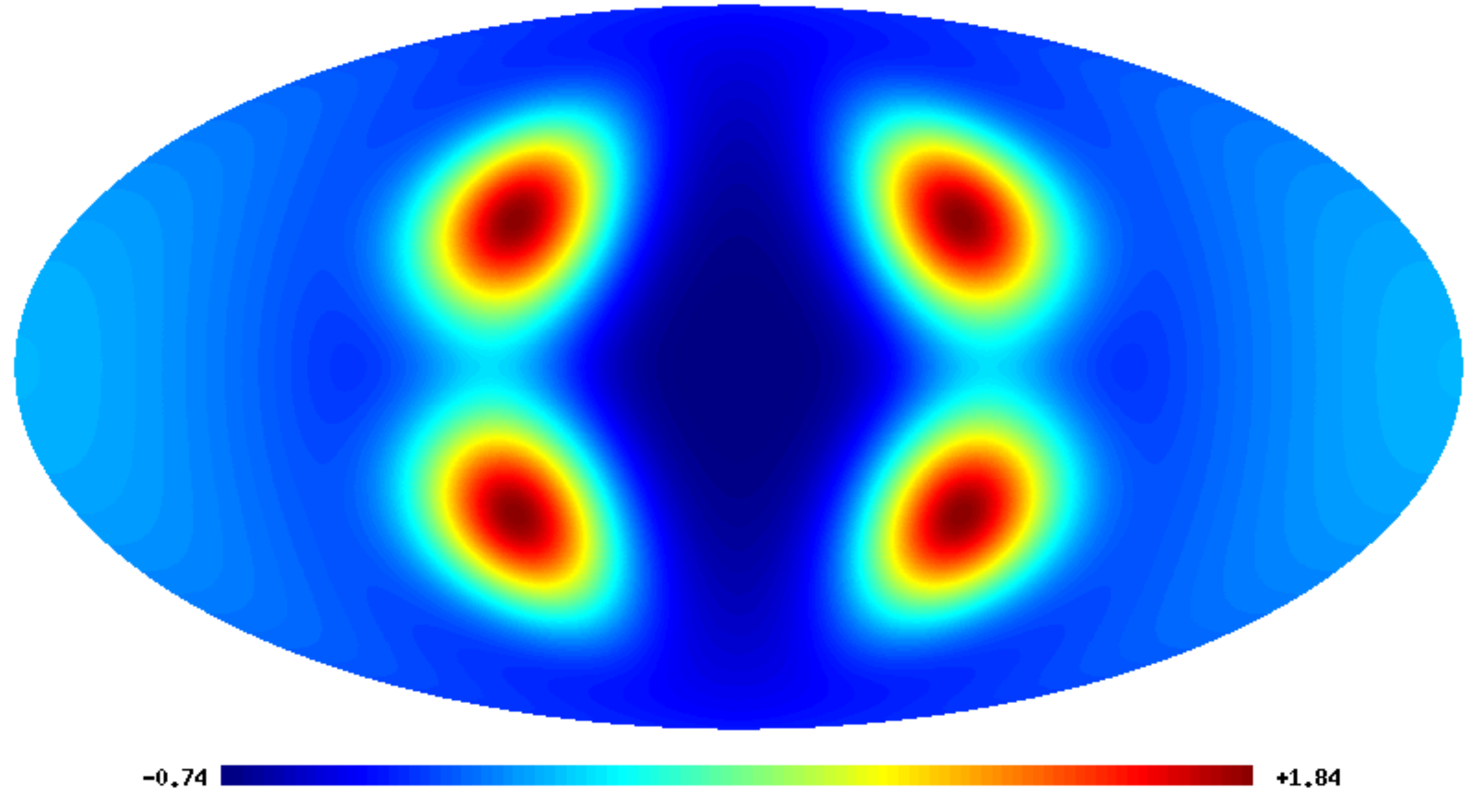}
\includegraphics[scale=.2]{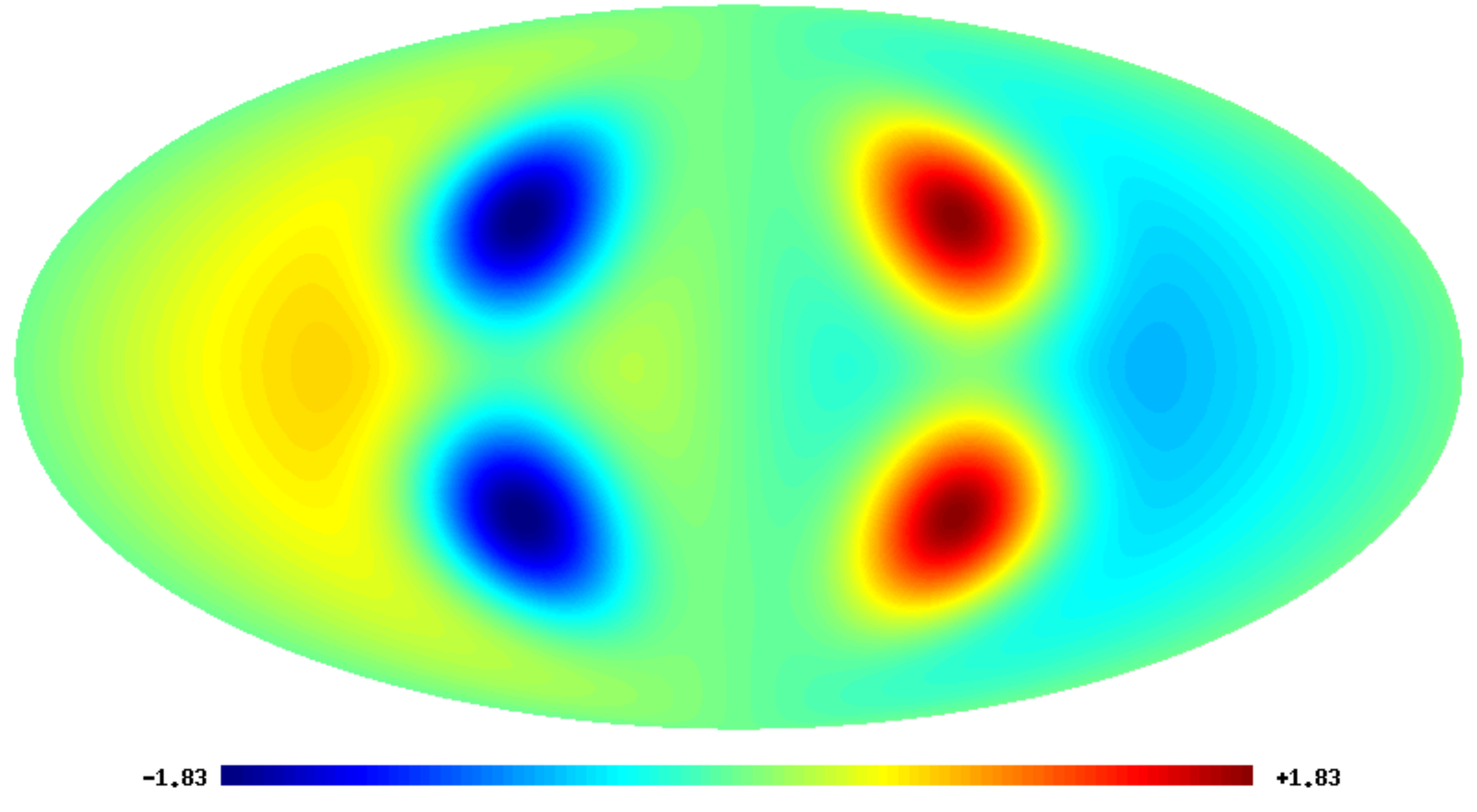}}
\centerline{\includegraphics[scale=.2]{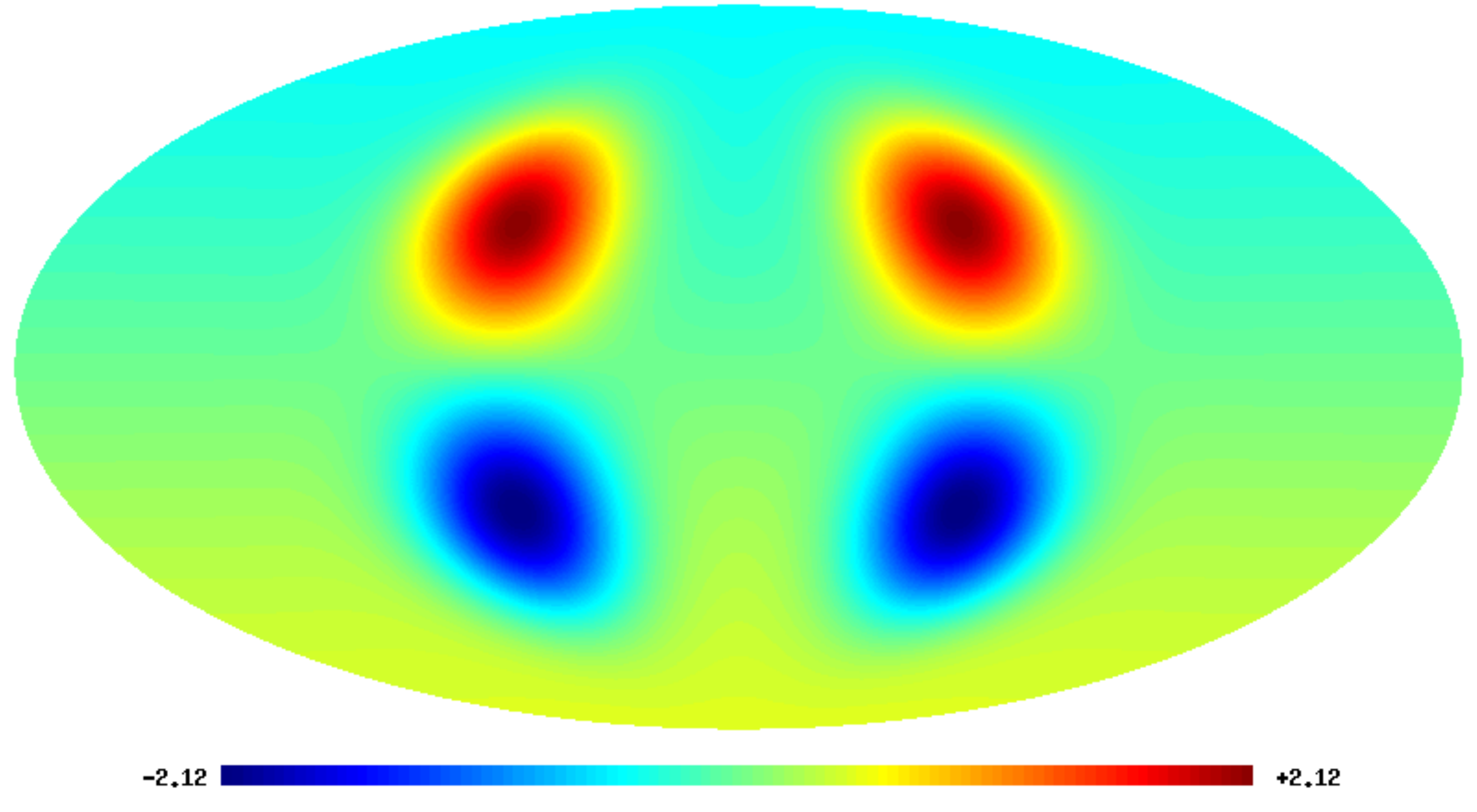}
\includegraphics[scale=.2]{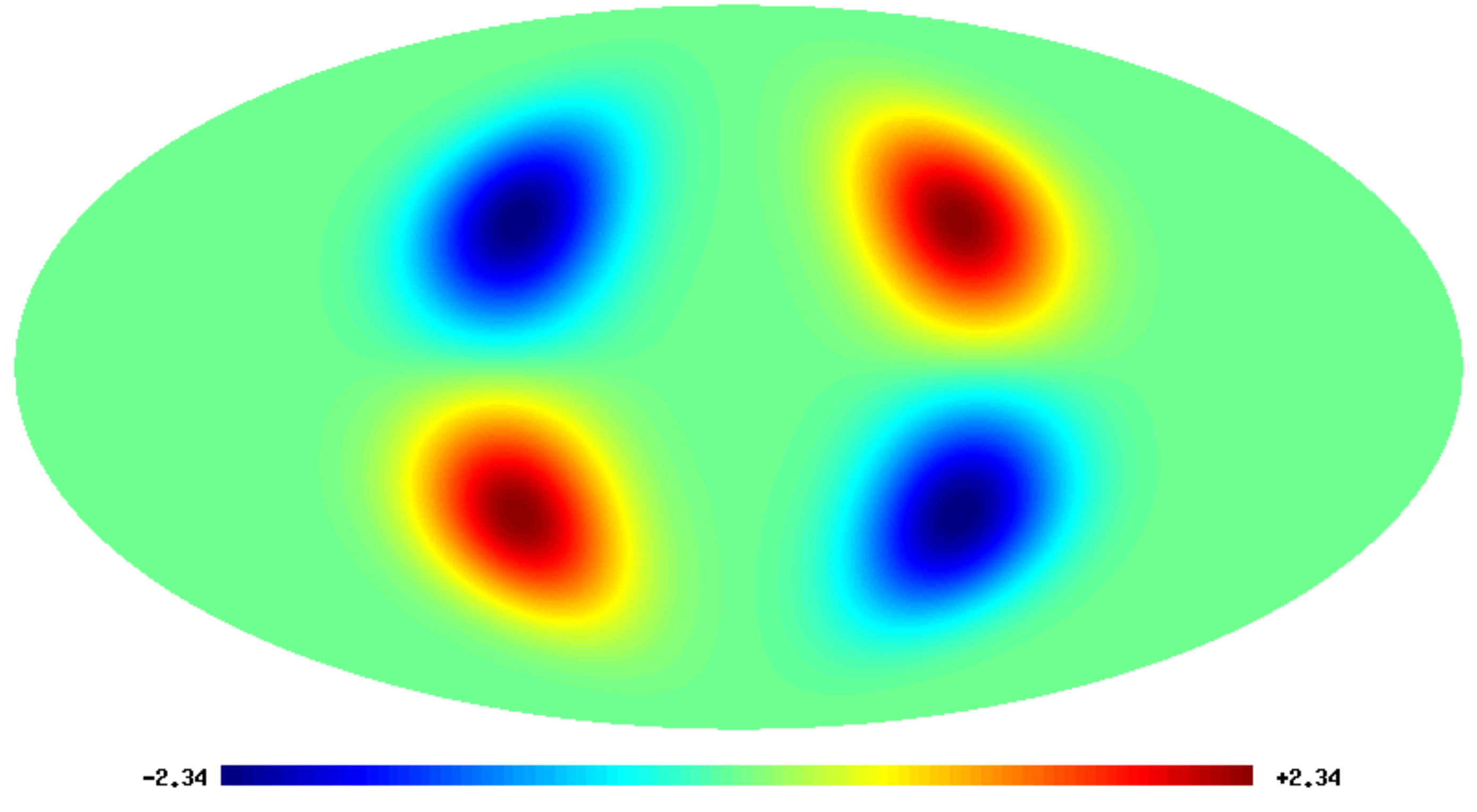}
}
  \caption{Top left. Classification of the symmetries of the CMB with respect to the antipodal points (S1), the Galactic plane (S2) and $\phi\rightarrow 2\pi-\phi$ (S3). Top right. The map of a signal with all multipoles and $m$-modes. Then, from the second from the top row (left) and down to the bottom right panel we show the maps for $l=even$, $l=odd$, $l+m=even$, $l+m=odd$, $l+m=even, \Im m=0$, $l+m=even,\Re e=0$, $l+m=odd,\Im m=0$, and $l+m=odd,\Re e \ m=0$. All in Galactic coordinates and Mollweide projection.}
 \label{fig1} }
For the CMB signals $\Delta T(\hat{\mathbf n})$ on the sphere, one can define the symmetric $ \Delta T^+(\hat{\mathbf n})$ and anti-symmetric $ \Delta T^-(\hat{\mathbf n})$ components:
\begin{eqnarray} 
\Delta T(\hat{\mathbf n})=\Delta T^+(\hat{\mathbf n})+\Delta T^-(\hat{\mathbf n}),
\end{eqnarray}
where all inversions have been given with respect to the origin of the polar system of coordinates, which is a standard basis for estimation of the $a_{l,m}$-coefficients according to Eq(\ref{eq1}), and 
\begin{eqnarray} 
\Delta T^+(\hat{\mathbf n})&=&\frac{\Delta T(\hat{\mathbf n})+\Delta T(-\hat{\mathbf n})}{2}%=\nonumber\\
=\sum_l\sum_{m=-l}^{l}a_{l,m}P^+(l)Y_{lm}(\hat{\mathbf n}),\nonumber\\
\Delta T^-(\hat{\mathbf n})&=&\frac{\Delta T(\hat{\mathbf n})-\Delta T(-\hat{\mathbf n})}{2}%=\nonumber\\
=\sum_l\sum_{m=-l}^{l}a_{l,m}P^-(l)
Y_{lm}(\hat{\mathbf n}),\nonumber\\
\end{eqnarray}
and $P^+(l)=\cos^2(\frac{\pi l}{2})$, $P^-(l)=\sin^2(\frac{\pi l}{2})$.

By definition, $\Delta T^+(\hat{\mathbf n})$ and $\Delta T^-(\hat{\mathbf n})$ are orthogonal in the sense that the averaged product over the whole sphere vanishes (i.e. $\langle \Delta T^+(\hat{\mathbf n})\Delta T^-(\hat{\mathbf n})\rangle=0$).
The power spectrum of $\Delta T(\hat{\mathbf n})$ is given by
\begin{eqnarray} 
 C(l)=\frac{1}{2l+1}\sum_m|a_{lm}|^2=C^+(l)+C^-(l),
\end{eqnarray}
where $C^+(l)=C(l)P^+(l)$ and $C^-(l)=C(l)P^-(l)$.
In other words, $C^+(l)$ and $C^-(l)$ are associated with the power spectrum of even and odd multipoles respectively.

The area near the Galactic plane contain a very powerful concentration of the Galactic diffuse foregrounds (synchrotron, free- free and dust emission), and is highly contaminated by the Galactic point-like sources. The Galactic plane is perpendicular to the z-axis and centered at $z=0$ in the Galactic coordinate system and, as we mentioned above, is the basis of a type of symmetry. The coordinate inversion from $\hat{\mathbf n}=(\theta,\phi)$ to $\overline{\hat{\mathbf n}}=(\pi-\theta,\phi)$ give us a symmetric $\Delta T_s(\hat{\mathbf n})=\Delta T_s(\overline{\hat{\mathbf n}})$ or an anti-symmetric $\Delta T_a(\hat{\mathbf n})=-\Delta T_a(\overline{\hat{\mathbf n}})$ signal with respect to that plane. Since $Y_{l,m}(\overline{\hat{\mathbf n}})=(-1)^{l+m}\,Y_{l,m}(\hat{\mathbf n})$, one can see that the symmetric signal $\Delta T_s$ corresponds to $l+m=even$ and $\Delta T_a$ corresponds to $l+m=odd$. Taking into account $\Delta T^+=\Delta T_s^+ +\Delta T_a^+$ and $\Delta T^-=\Delta T_s^- +\Delta T_a^-$, we can see that the most symmetric part of the signal $\Delta T$ (with respect to the origin and the plane at $z=0$) corresponds to $\Delta T_s^+$ with $l=even$, $l+m=even$, and the most anti-symmetric part is $\Delta T^-_a$ with $l=odd$ and $l+m=odd$. 
We have summarized all the mentioned symmetries with the illustration in Fig.\ref{fig1}. Note that all $a_{l,m}$ coefficients were found from the polar system of coordinates, centered at the Galactic center. Then, by selecting different multipoles (for instance, only even $l$, or only odd), we plot the corresponding maps in Galactic coordinates and in Mollweide projection.

\section{The ILC 7 octupole. General properties.}

As an illustration of the different kind of symmetries in the CMB sky, we plot the octupole component of the ILC7 map in Fig.\ref{fig1a}-\ref{fig2}. Note that the octupole is the most powerful anti-symmetric component with respect to the inversion $\hat{\mathbf n}\rightarrow -\hat{\mathbf n}$ on the sky. 
For the octupole (i.e. $l=3$), even and odd $m$ number corresponds to odd and even values of $l+m$.
From Eq. \ref{eq1}, one can see that symmetry $\phi\rightarrow 2\pi-\phi$, $\theta=const$ for even and odd $m$ is associated with real and imaginary part of the spherical harmonic coefficients (i.e. $\Re[a_{l,m}]$ and $\Im[a_{l,m}]$).\\
\FIGURE{
   \hbox{\centerline{\includegraphics[scale=.22]{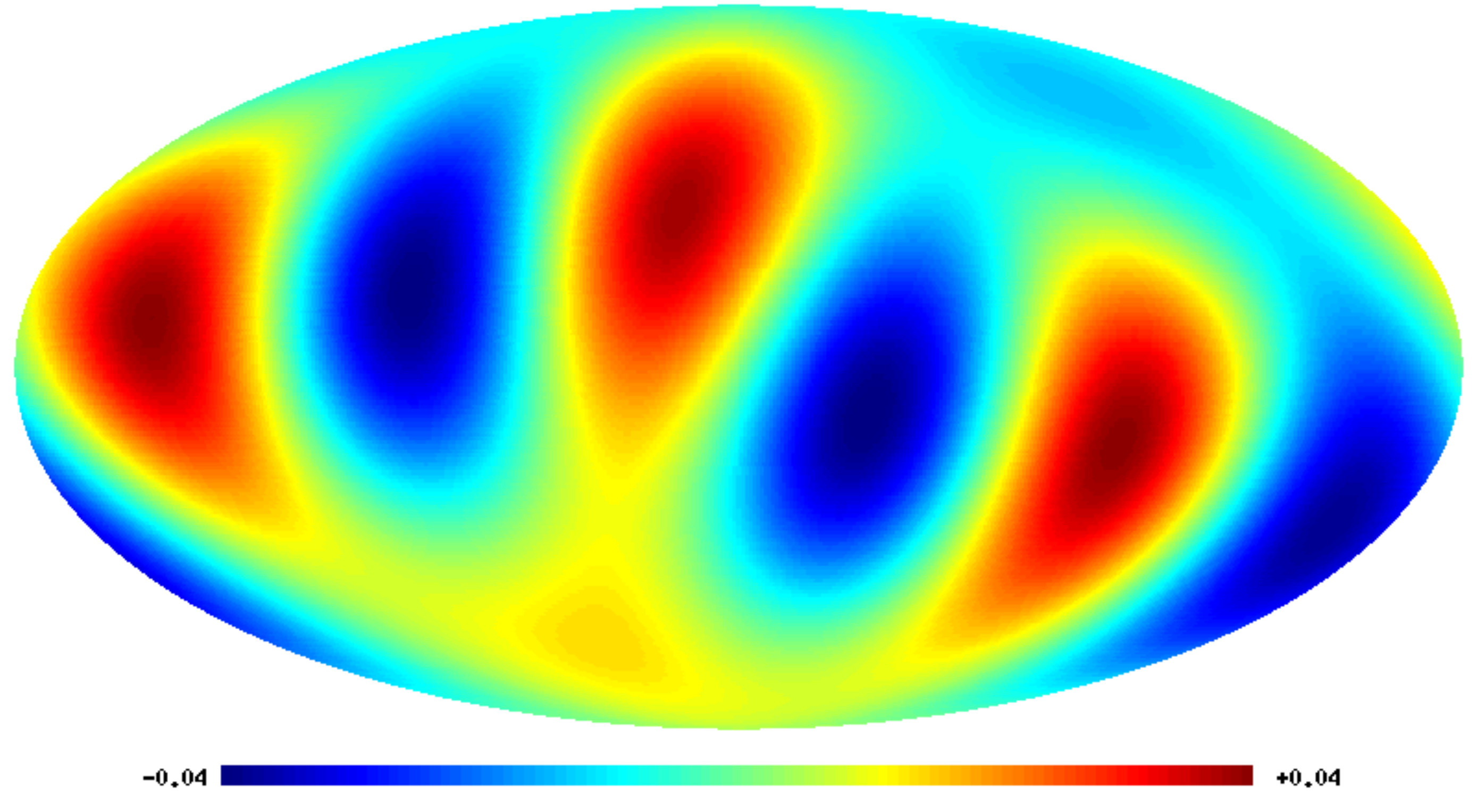}
\includegraphics[scale=.22]{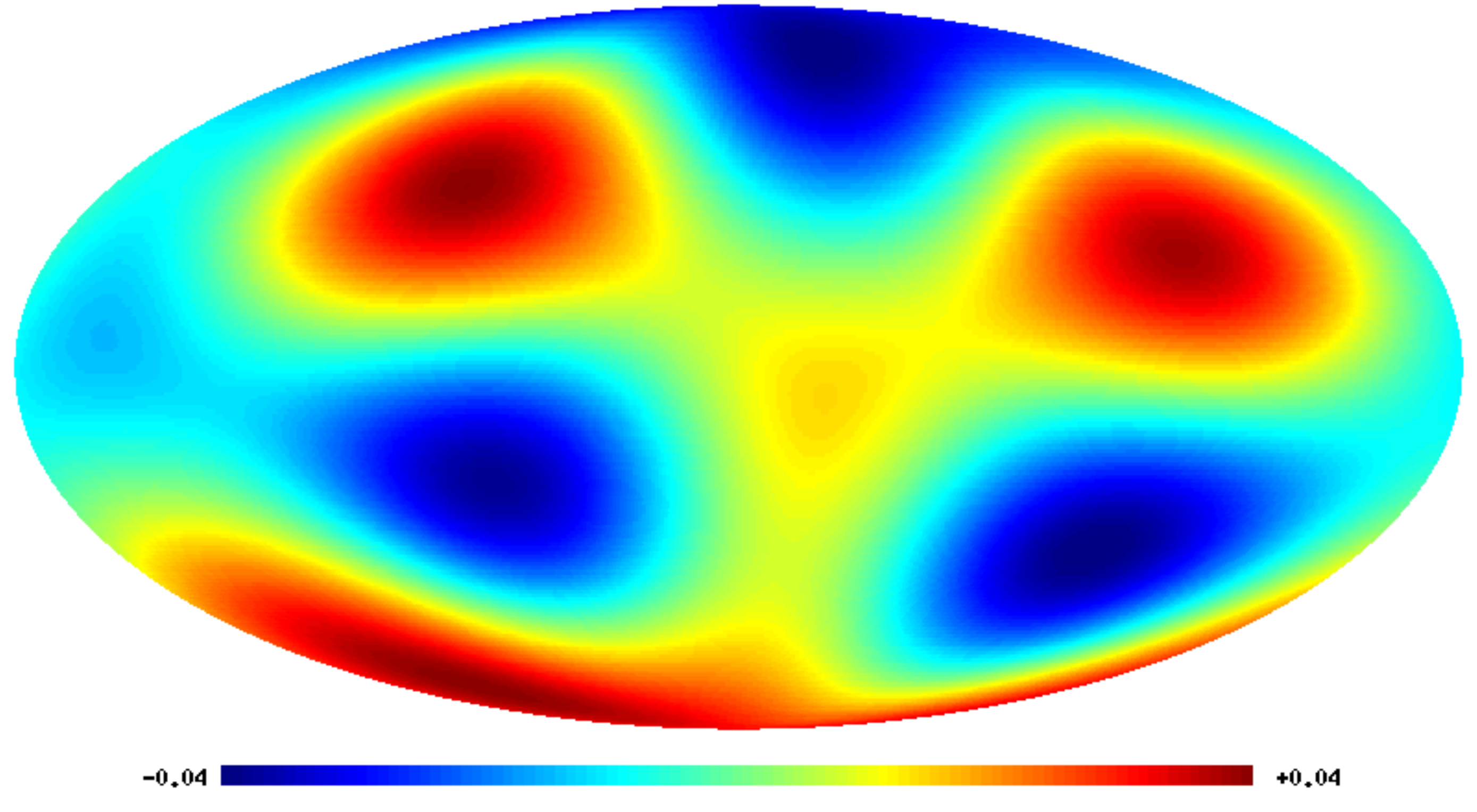}}}
\caption{The map of the octupole in Galactic (left) and ecliptic (right) coordinates.}
\label{fig1a}}

\FIGURE{
   \hbox{\centerline{\includegraphics[scale=.22]{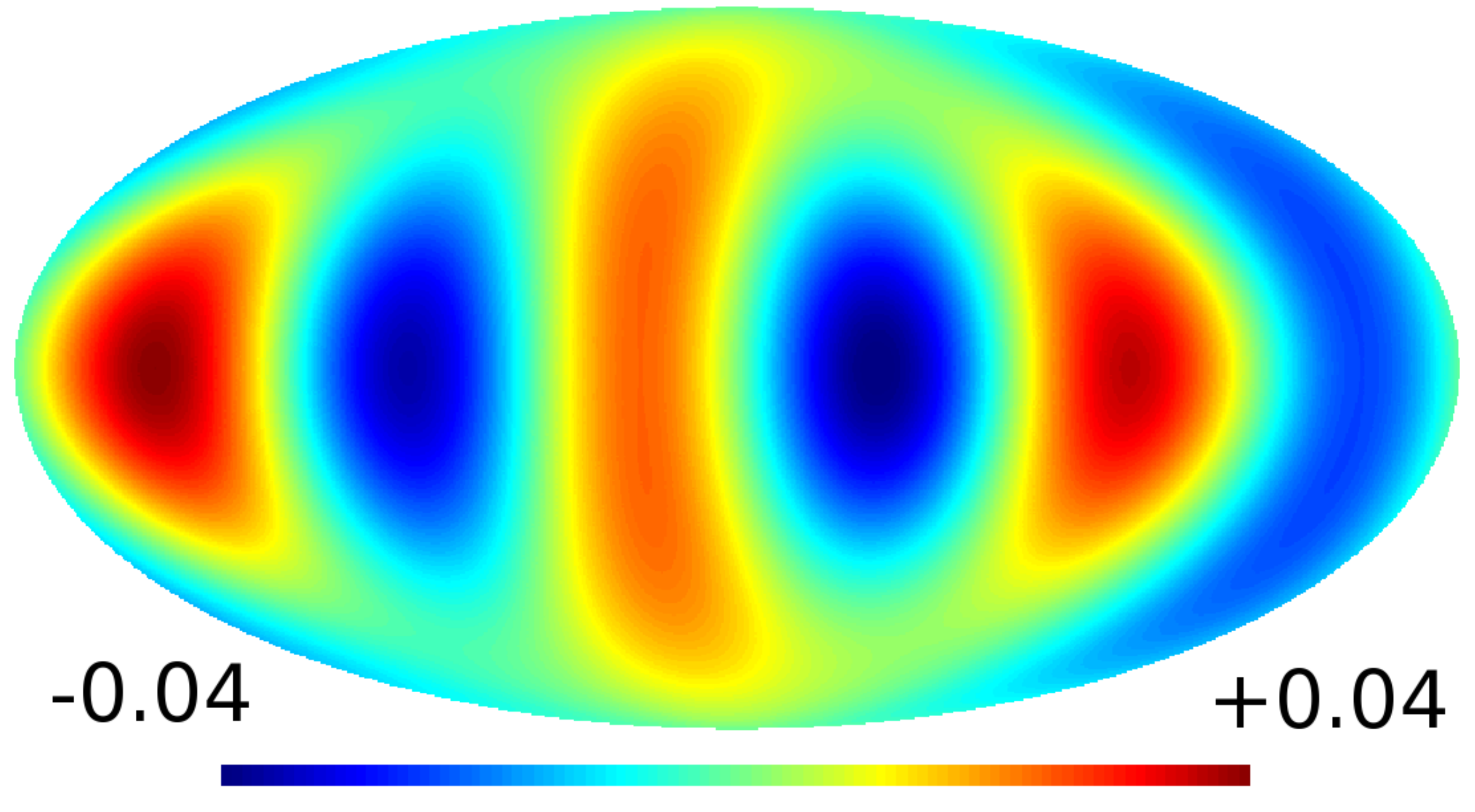}
\includegraphics[scale=.22]{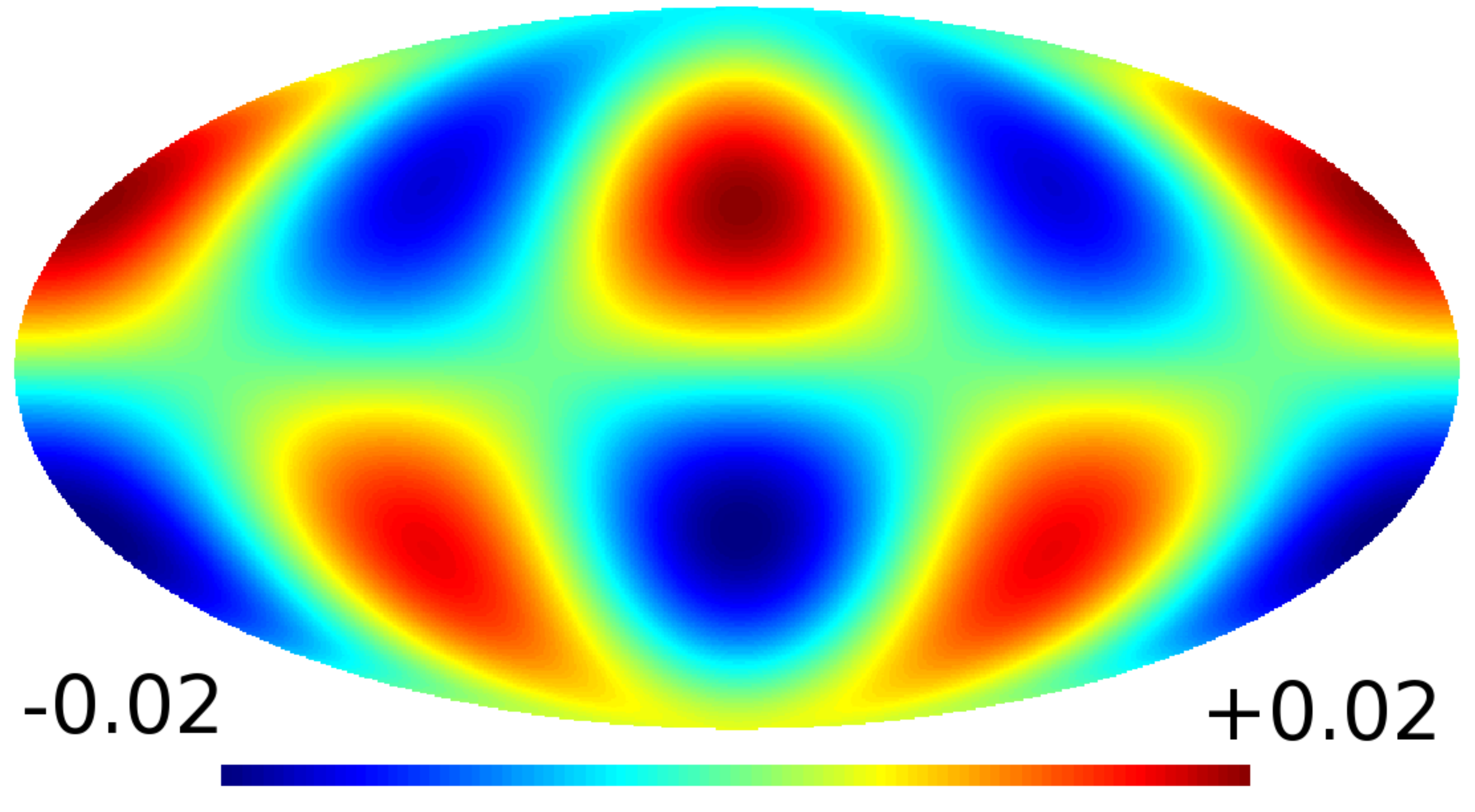}}}
   \hbox{\centerline{\includegraphics[scale=.22]{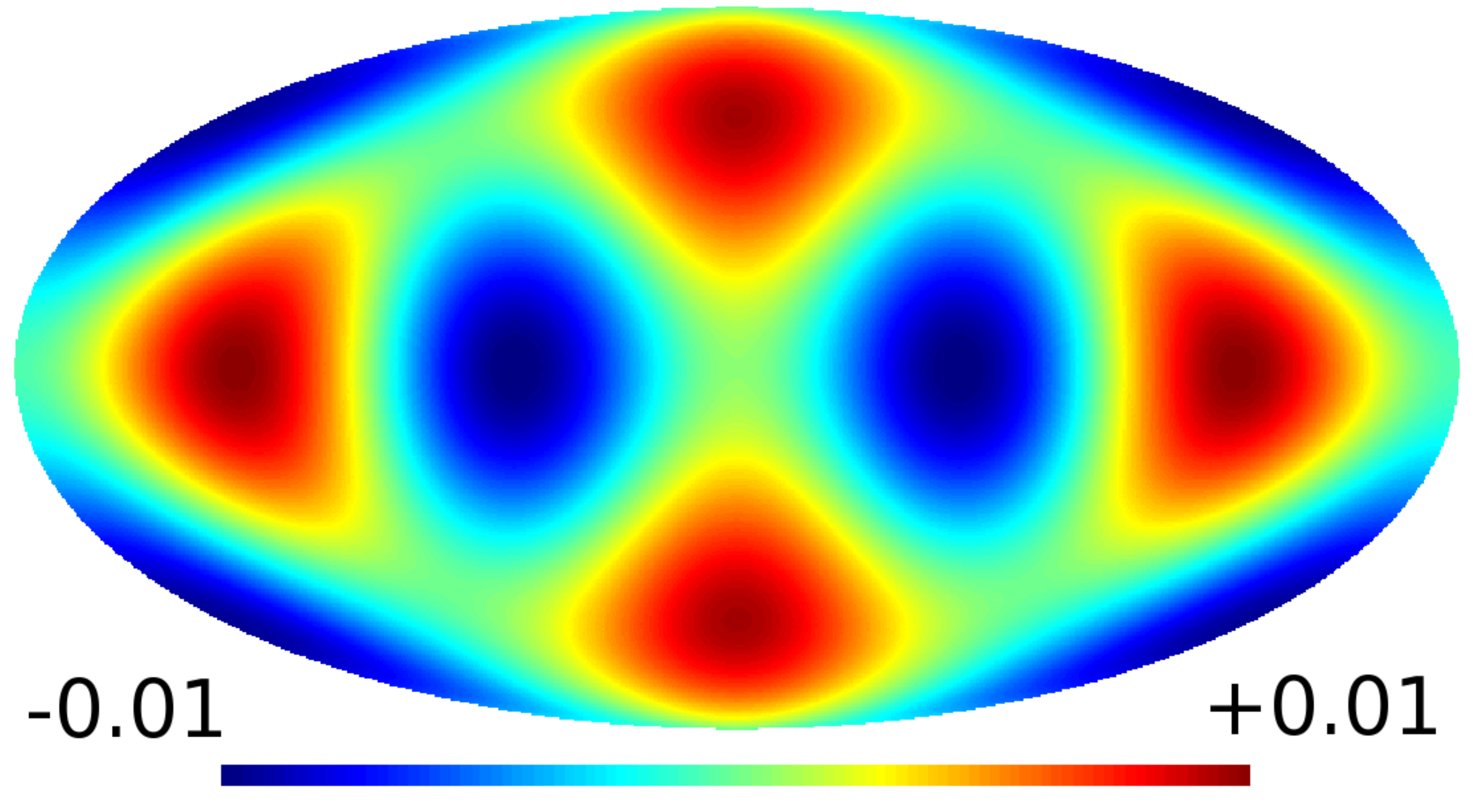}
\includegraphics[scale=.22]{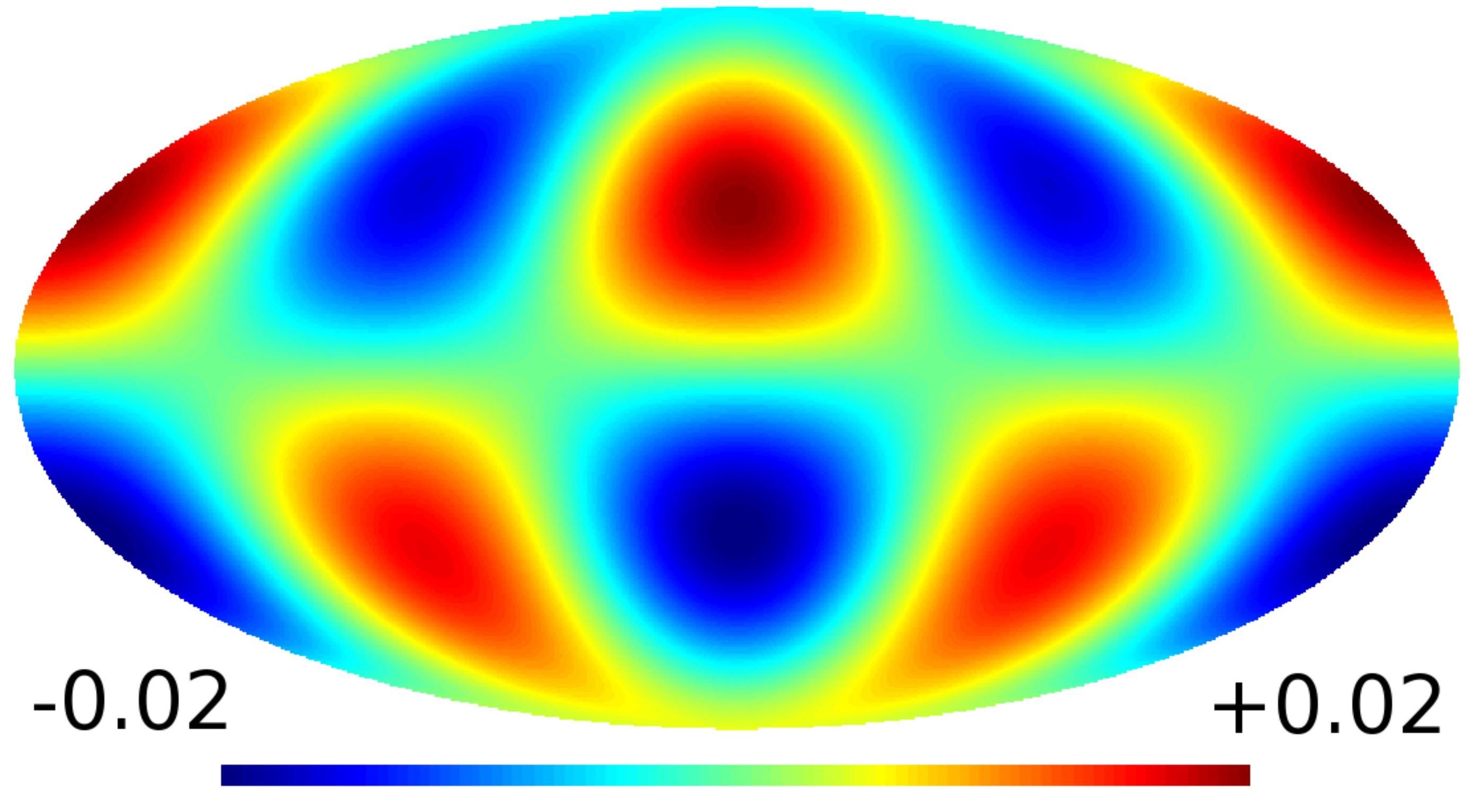}}}
   \hbox{\centerline{\includegraphics[scale=.22]{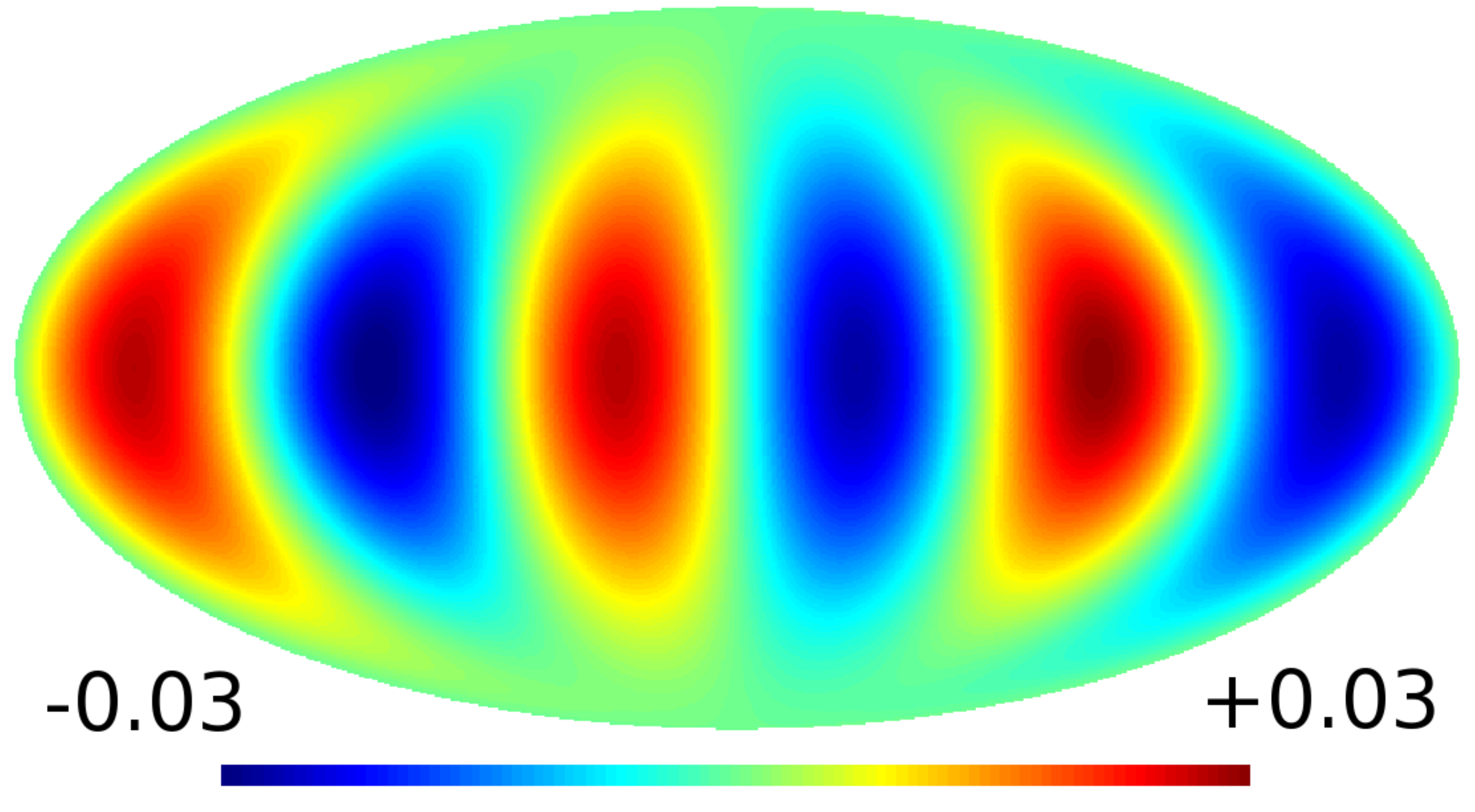}
\includegraphics[scale=.22]{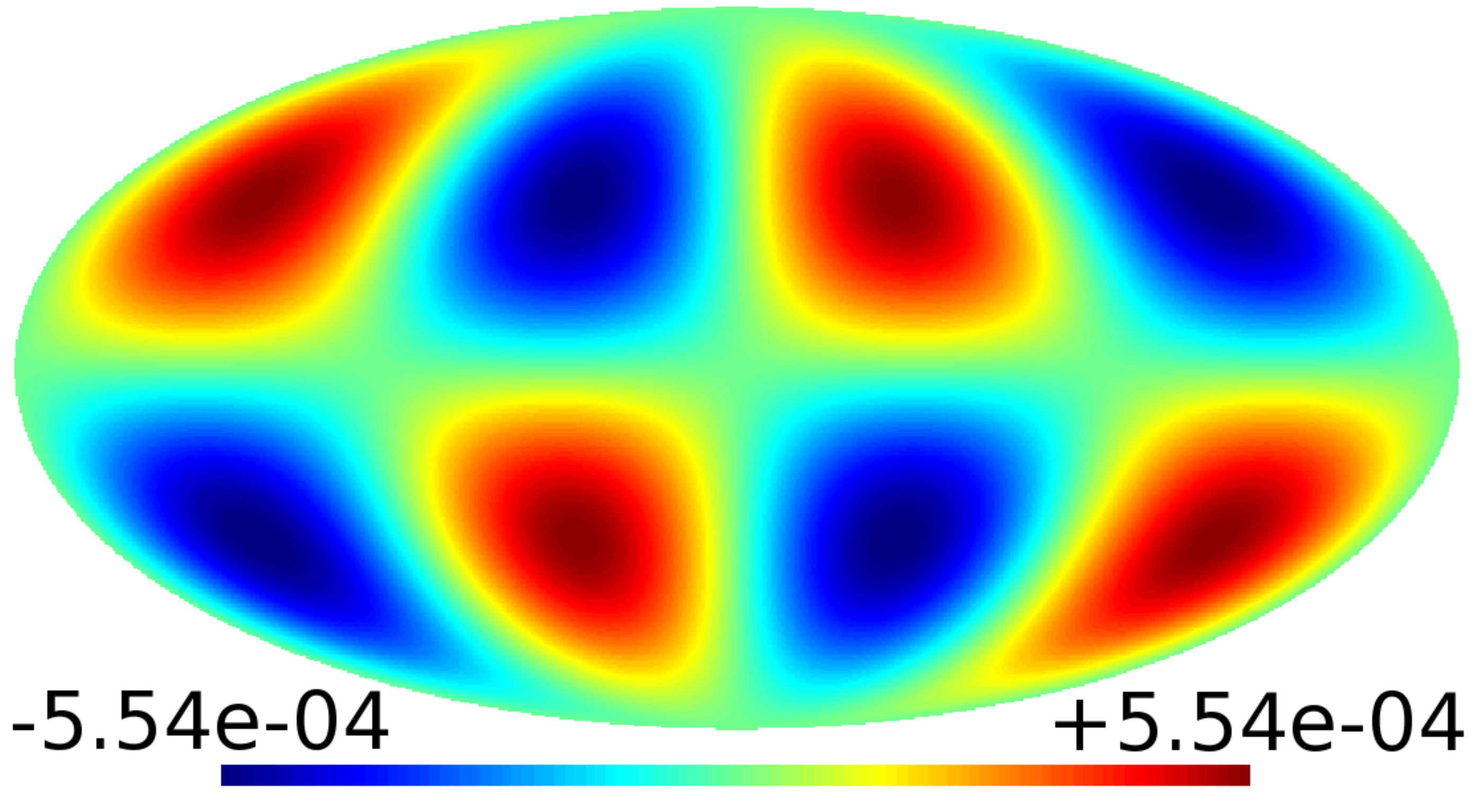}}}
\caption{Components of the ILC7 octupole: On the left and the right side, we show the octupole components, where only $a_{lm}$ of even and odd $l+m$ are retained. At the same time, we have retained both real and imaginary part of $a_{lm}$ (top), real part of $a_{lm}$ (middle) and imaginary part of $a_{lm}$ (bottom).}
\label{fig2}}
Coming back to Eq(\ref{eq1}, we can see that different symmetries or asymmetries of the octupole component are connected with
corresponding components of $a_{3,m}$ coefficients, which now play a role of weighting coefficients for different symmetric or asymmetric terms (see table \ref{table1a}). For a statistically homogeneous and isotropic random Gaussian field the distribution of the phases of $a_{3,m}$ for $m=0,1,2,3$ is uniform in the interval $[0,2\pi]$, while the amplitudes $|a_{l,m}|$ are distributed according to the Rayleigh probability density function.
\TABLE{
\caption{ $\Re e$ and $\Im m$ parts (in $mK$) of the multipole coefficients $a_{3,m}$ for the ILC 7. The corresponding power
for $l+m=even$ is $C_{even}=2.02{10}^{-4} mK^2$, and for $l+m=odd$ we have $C_{odd}=7.51{10}^{-5} mK^2$}
\begin{tabular}{ccc}
\hline 
 $l=3,m$    &     $\Re e (a_{3,m})$ &$ \Im m(a_{3,m})$\\
\hline
$m=0$ & $ -6.479e-03 $& $0 $\\ 
$m=1$ & $-1.219e-02  $ & $ 2.0265e-03$ \\ 
$m=2$ & $2.199e-02 $ & $5.907e-04 $  \\ 
$m=3$ & $ -1.171e-02$ & $3.355e-02 $ \\
\hline
\end{tabular}
\label{table1a} }

\subsection{Symmetry estimators as an example of Cauchy distribution.}

We now investigate the $a_{l,m}$'s of the octupole. It is clear, that the value of $a_{l,m}$ that best conforms to the criteria for asymmetry is $a_{3,2}$, where $l=odd$ and $l+m=odd$. Naturally very many values of $a_{l,m}$ meet this criteria, but $a_{3,2}$ is the one with most power, as mentioned above. We now wish to test the ratio between the $a_{3,2}$-multipole, and its more symmetric neighbors $a_{3,1}$ and $a_{3,3}$ (where $l+m=even$), for both the real and imaginary value. To do this, we introduce the following estimators:
\begin{eqnarray}
%\nonumber
\alpha_{1}=\frac{\Im(a_{3,1})}{\Im(a_{3,2})}, \qquad \alpha_{3}=\frac{\Im(a_{3,3})}{\Im(a_{3,2})},\qquad 
\beta_{1}=\frac{\Re(a_{3,1})}{\Re(a_{3,2})}, \qquad \beta_{3}=\frac{\Re(a_{3,3})}{\Re(a_{3,2})}
\label{alpha}
\end{eqnarray}

If all the real and imaginary parts of the estimators follow a random Gaussian process, the parameters $\alpha_{1},\alpha_{3},\beta_{1}$ and $\beta_{3}$ follow a Cauchy distribution function. This particular type of random process is characterized by the probability density function:
\begin{eqnarray}
 f(x)=\frac{A}{(x-x_0)^2 +\gamma^2}
\label{ch}
\end{eqnarray}
 where $A$ is the normalization constant, $x_0$ is the location parameter, and $\gamma$ is the probable error.
The Cauchy distribution is one of the examples of a random distribution without mean value, variance or higher order moments. However the probability $P(x>X)$ to get some corresponding values of $x>X$ is still defined, and it is given by the integral in Eq(\ref{prob}).\\
For the random Gaussian field, when nominators and denominators in Eq(\ref{alpha}) are normalized to $\sqrt{0.5C(l)}$, we have $A=1/\pi$, $x_0=0$ and $\gamma=1$, and thus the result in Eq(\ref{prob}) can be simplified significantly.
\begin{eqnarray}
 P(x>X)=A\int_X^{\infty}\frac{dx}{(x-x_0)^2+\gamma^2}=\frac{A}{\gamma}\left(\frac{\pi}{2}-tan^{-1}\left(\frac{X-x_0}{\gamma}\right)\right)\simeq\frac{1}{\pi X},\hspace{0.1cm} X\gg 1
\label{prob}
\end{eqnarray}

\FIGURE{
    \centerline{\includegraphics[scale=0.6]{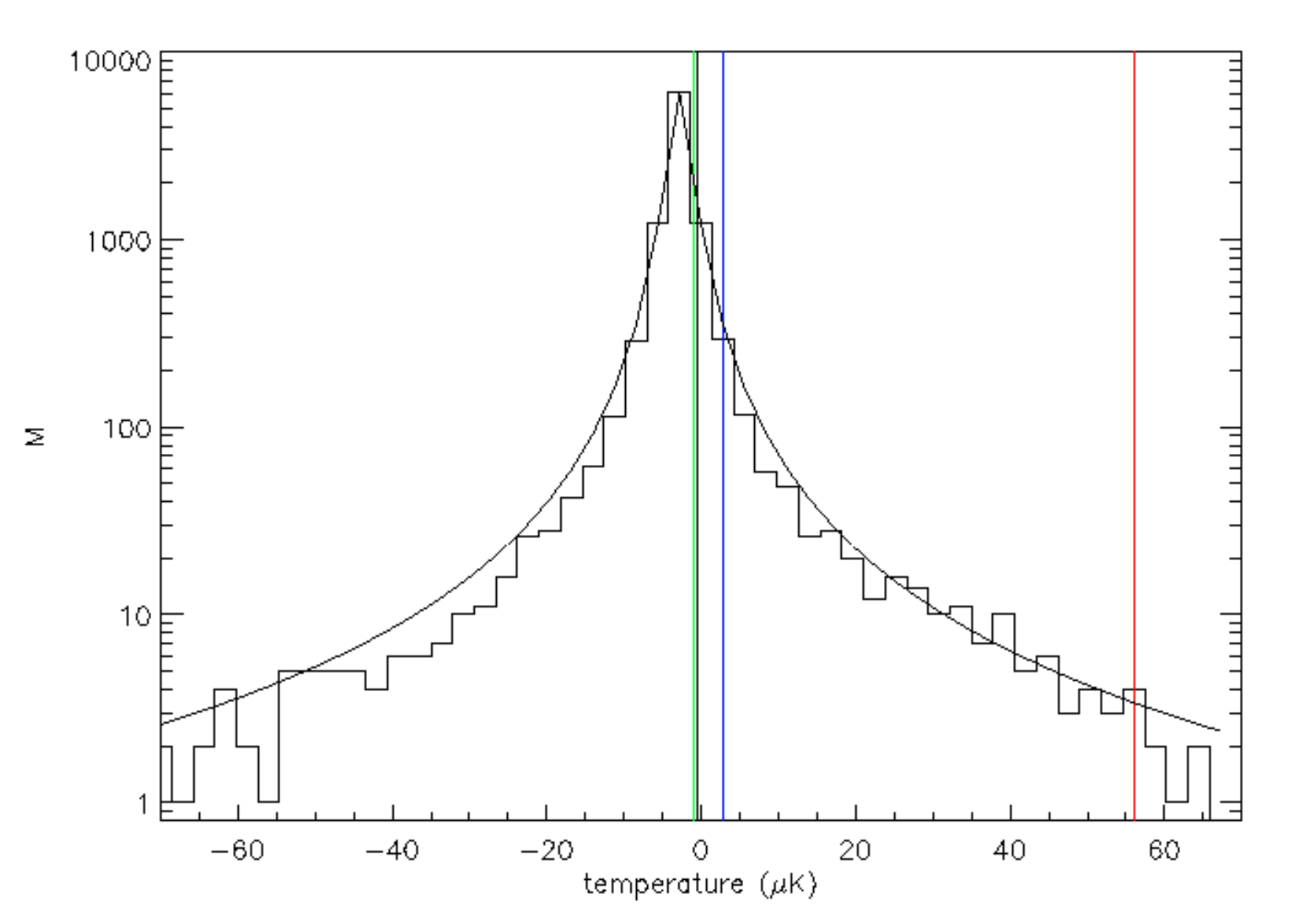}}
%                \includegraphics[scale=0.4]{alpha3_10000.pdf}}
 %   \centerline{\includegraphics[scale=0.4]{beta1_10000.pdf}
 %               \includegraphics[scale=0.4]{beta3_10000.pdf}}
    \caption{Value of the estimators from eq. \ref{alpha}, compared to a histogram of 10.000 Monte Carlo simulations. The blue line corresponds to $\alpha_{1}$, the red line is $\alpha_{3}$. The green line show $\beta_{1}$ the vertical black line (somewhat obscured by the green) corresponds to $\beta_{3}$. Note that the y-axis is in log. The smoothed black line corresponds to the Cauchy distribution.}
    \label{histograms} }

\TABLE{
\caption{Table of $P(x>X)$ for $X=|\alpha_{1}|,|\alpha_{3}|,|\beta_{1}|,|\beta_{3}|$ for the WMAP octupole and the 10.000 Monte Carlo simulations (MC) in comparison with $P(x>X)$ from Eq(\ref{prob})(Th). }
\begin{tabular}{cccc}
\hline 
     &    WMAP7 & MC & Th\\
\hline
$|\alpha_{1}|$  & $3.43$ & $0.093$& $0.093$\\ 
$|\alpha_{3}|$ & $56.8$ & $0.0058$ & $0.0056$ \\ 
$|\beta_{1}|$ & $0.55$ & $0.66$ & $0.65$ \\ 
$|\beta_{3}|$ & $0.53$ & $0.65$ & $0.64$ \\ 
\hline
\end{tabular}
\label{table1} }
We have performed the test on the WMAP ILC 7 year data, and compared with 10.000 Monte Carlo simulations of the CMB sky. Note that we run numerical simulations in order to assess the level of fluctuations for finite sample and $X\ll 1$. One more remark is connected with the properties of the Cauchy distribution, given by estimators from Eq(\ref{alpha}). Since all the estimators are based on the ratios of (potentially) Gaussian variables, the big values of them seems to be very unlikely, as well as small values $X\rightarrow 0$. The point is that inversions $|\alpha_i|\rightarrow |\alpha_i|^{-1}$ and $|\beta_i|\rightarrow |\beta_i|^{-1}$ transform small values of the parameters to big one, leaving the Cauchy distribution in form of Eq(\ref{ch}) unchangeable. The results are plotted in fig. \ref{histograms}, and in table \ref{table1}.

We see a big deviation for the $\alpha_{3}$ parameter, with only 58 out of 10.000 simulations having a larger value, corresponding to $0.58\%$, while from Eq(\ref{prob}) we have corresponding probability $P(x>X=56.8)=0.0056$. For the $\alpha_{1}$ parameter, we have $9.4\%$, and for the two $\beta$ parameters, the deviation is not significant. Thus it seems, that the imaginary part of the $a_{3,2}$ multipole carries little weight, especially compared to the $a_{3,3}$ multipole. It is clear, that the octupole exhibit a symmetric behavior, and contributions from the foreground seems to be one of the most likely explanations. However, in the next section we will show that this problem could be more complicated, than the very natural assumption, that only the foreground component is responsible for the peculiar symmetry of the WMAP octupole.

\subsection{The WMAP 7 octupole. The phases.}
As it was pointed out in Section 2, the S3 symmetry reflect directly the properties of real and imaginary parts of the $a_{l,m}$ coefficients. In application to the octupole component, the corresponding estimators $\alpha_1,\alpha_3,\beta_1$ and $\beta_3$ were designed to evaluate different symmetries or asymmetries of the WMAP octupole. However, for a given value of $m$ the ratio between imaginary and real parts of the $a_{l,m}$ is nothing, but the phase of the corresponding $l,m$ harmonic, defined as
\begin{eqnarray}
 \psi_{l,m}=tan^{-1}\left(\frac{\Im m ~a_{l,m}}{\Re e ~a_{l,m}}\right)
\label{phase}
\end{eqnarray}
According to \cite{Chiang}, \cite{naselsky}, the phases of the $a_{l,m}$ coefficients reflect directly the morphology of the $\Delta T$-map, and for the octupole component they are connected with $\alpha_1,\alpha_3,\beta_1$ and $\beta_3$ as follows:
\begin{eqnarray}
 \alpha_1=\frac{\tan(\psi_{3,1})}{\tan(\psi_{3,2})}\beta_1,\hspace{0.3cm} \alpha_3=\frac{\tan(\psi_{3,3})}{\tan(\psi_{3,2})}\beta_3
\label{alpha-ph}
\end{eqnarray}

The analysis of the phases is very useful for investigating possible cross-correlations between two or more signals, $a_{l,m}$ and $f_{l,m}$, where $f_{l,m}$ corresponds to the foreground component (residuals of the Galactic diffuse foreground and extra-galactic point-like sources), for instance, the WMAP V or W bands. At the same time, the multipole vectors approach \cite{Copi1},\cite{Copi5} clearly detect the coupling between the WMAP quadrupole and octupole and kinematic (non-cosmological ) dipole. This result was never tested before by analysis of phases of the octupole component, as presented below. In Fig.\ref{phas} we show the phases of the WMAP ILC7 octupole and NILC5 octupole from \cite{nilc} versus the phases of the kinematic dipole.
 
\FIGURE{
    \centerline{\includegraphics[width=0.5\linewidth]{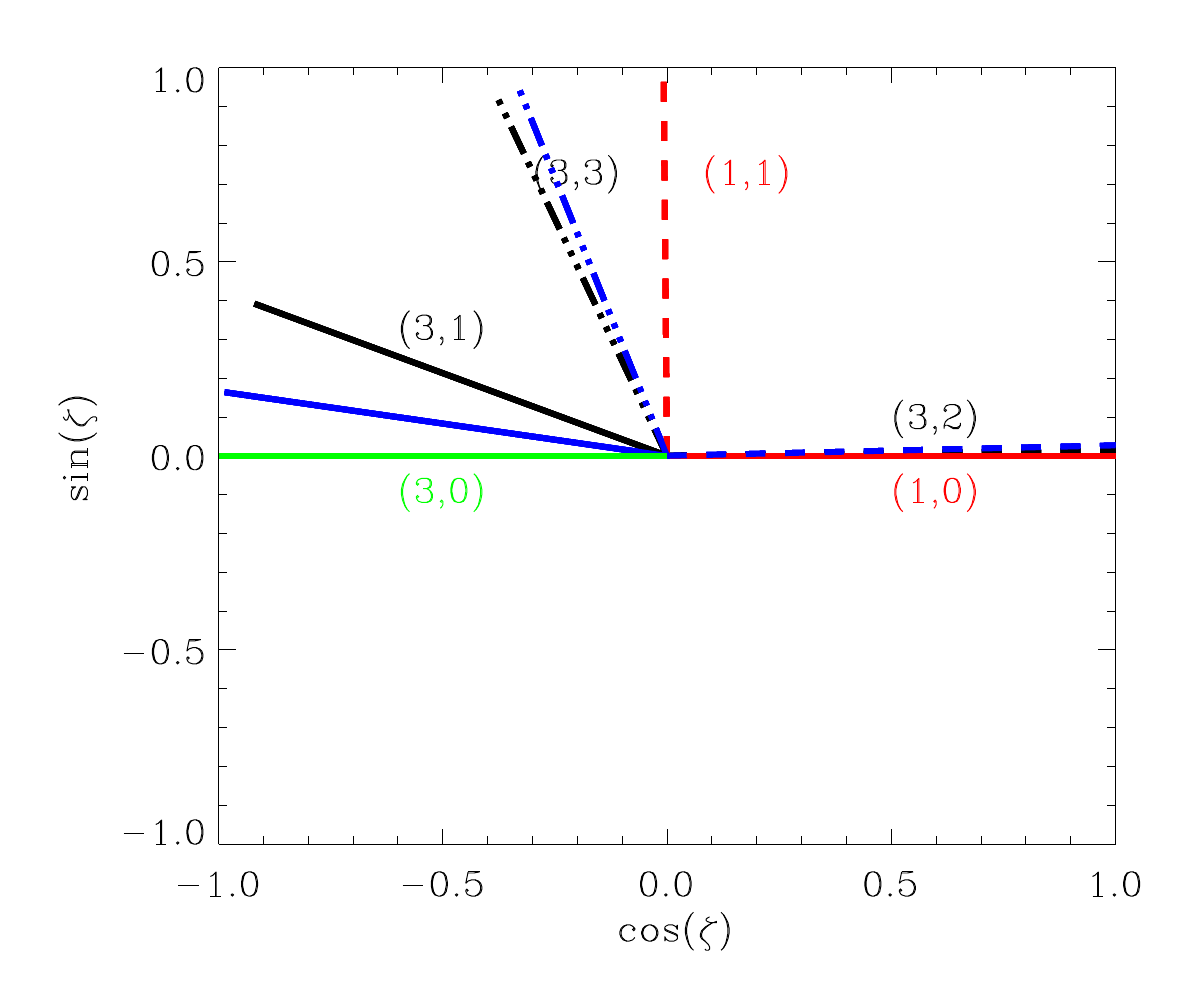}}
    \caption{The phases for NILC5 (black) and ILC7 (blue) octupole. The solid line is for $\psi_{3,1}=2.976$ rad, the dash line is for $\psi_{3,2}=0.0268$ rad. and the dash-dotted line is for $\psi_{3,3}=1.907$ rad. The red solid line is for dipole $\Psi^d_{1,0}=0$, and the red dash line is for $\Psi^d_{1,1}=\pi/2$ phases.}
    \label{phas} }

Due to periodicity of the phases within the interval $-\pi,\pi$, the estimators of cross-correlations between different signals are based on the trigonometric moments $G_{l,m}^{l',m'}= \cos(\psi_{l,m}-\Psi_{l',m'})$(see \cite{naselsky} for review). Thus for the WMAP 7 octupole and the phases of the kinematic dipole, the corresponding estimators are :
$$
 G_{3,1}^{1,0}=-0.986,\hspace{0.2cm}G_{3,2}^{1,0}=0.99964, \hspace{0.2cm} G_{3,3}^{1,1} =0.9440.
$$
One can see that the phase correlations of the ILC 7 octupole and the kinematic dipole is especially strong for $3,2$ component, practically at the same level of significance, as for $\alpha_3$ test. Does it mean, that the major source of peculiar symmetry of the octupole is connected to the residuals of the dipole substructure? Answering that question, we would like to stress, that in Fig.\ref{phas} the phases of the NILC5 octupole are slightly different from the corresponding phases of the ILC 7 octupole. The NILC approach differs from the WMAP ILC method by different evaluation of the CMB-foreground coupling. Taking under consideration the templates derived by the WMAP team for the synchrotron, free-free, and dust emission, we have synthesized the sum of all these components for the V band, in order to obtain the phases $\Phi_{3,m}$ of the octupole component, listed below: 

$$
\Phi_{3,0}=0,\hspace{0.2cm}\Phi_{3,1}=-0.0342,\hspace{0.2cm}\Phi_{3,2}=1.5126,\hspace{0.2cm}\Phi_{3,3}=-2.47629
$$
and further: $\cos(\psi_{3,1}-\Phi_{3,1})= -0.991$, $\cos(\psi_{3,2}-\Phi_{3,2})=0.0849$ and $\sin(\psi_{3,2}-\Phi_{3,2})=0.9963$, and $\cos(\psi_{3,3}-\Phi_{3,3})=-0.323$. 
Thus, from the analysis of the $\psi_{l,m}$ and $\Phi_{l,m}$ cross-correlations, we may conclude that $3,1$ component of the octupole anti-correlate with $3,1$ phase of the foreground, and $3,2$ component of the octupole is orthogonal to the $3,2$ component of the foreground. This is why, we believe, the simplest assumption that only the foreground or only the kinematic dipole are responsible for the abnormal symmetry of the ILC 7 octupole, does not reproduce the whole picture of the possible contamination of the primordial signal.

Lets illustrate this idea by an analysis of the following model of contamination. Our assumption is, that the WMAP octupole component contains the primordial signal $c_{3,m}$, contaminated by the residuals of the foregrounds $f_{3,m}$ and residuals of the kinematic dipole as follows:

\begin{eqnarray}
 a_{3,m}&=&c_{3,m}+\sum_{m'=0}^1\hat{L}(3,m|1,m')d_{1,m'} +\mu f_{3,m},\hspace{0.2cm} m=1,2; 
\label{dip}
\end{eqnarray}
where: $\hat{L}(3,m|1,m')$ is the linear shift operator, which takes the dipole components to $3,2$ and $3,3$ components of the octupole, and $\mu$ is the ``ILC-residuals of the foreground'' coupling parameter. For the imaginary part of $m=2$ component of the octupole from Eq(\ref{dip}) we get:
\begin{eqnarray}
 |a_{3,2}|\sin(\psi_{3,2})=|c_{3,2}|\sin(\xi_{3,2})+\Im m\left(\sum_{m'=0}^1\hat{L}(3,m|1,m')d_{1,m'}\right)+\mu|f_{3,2}|\sin(\Phi_{3,2}),\nonumber\\
\label{dip1}
\end{eqnarray}%\nonumber\\
 Since $\sin(\psi_{3,2})\simeq 0$, from Eq(\ref{dip1}) one can get:

\begin{eqnarray}
|c_{3,2}|\sin(\xi_{3,2})\simeq -\Im m \left(\sum_{m'=0}^1\hat{L}(3,m|1,m')d_{1,m'}\right)-\mu|f_{3,2}|\sin(\Phi_{3,2})
\label{dip2}
\end{eqnarray}
Thus, in framework of linear model of the ILC contamination, the primordial component of the octupole is fully determined by the residuals of the foregrounds and kinematic dipole. This result clearly illustrate the importance of investigation of the symmetries for different components of the ILC signal, if we are interested in a more accurate estimation of the statistical properties of the CMB signal, especially for high multipoles $l$.

\section{S2 -pathfinder of peculiar multipoles.}
In the previous section we have shown that the implementation of the symmetry test is very informative in application to the analysis of the anomalies of the low multipole tail of the CMB power spectrum. However, for high multipoles $l\gg 10$, the number of possible permutations of real and imaginary parts of the corresponding $a_{l,m}$ coefficients grows very rapidly. This is why for this range of multipoles it would be essential to propose the quick search test, based on the above mentioned symmetries $S1-S2$. As the basis of this test, we compute the power spectrum, using the real part and the imaginary part of $a_{lm}$ respectively, where the sum only includes either even or odd values of $l+m$.
\begin{eqnarray}
\label{cl+real}
C(l)&=&D(l)_{Re}^{+}+ D(l)_{Re}^{-}+D(l)_{I m}^{+}+ D(l)_{Im}^{-}, \nonumber\\
D(l)_{Re}^{\pm} &=& \frac{1}{2l + 1} \left[ a^2_{l0} P^{\pm}(l) + 2\sum_{m=1}^{l} \Re^2 (a_{lm})G^{\pm}(l,m) \right], \nonumber\\
D(l)_{Im}^{\pm} &=& \frac{2}{2l + 1} \sum_{m=1}^{l} \Im^2 (a_{lm})G^{\pm}(l,m), 
\end{eqnarray}
where $G^{+}(l,m)=\cos^2(\frac{\pi (l+m)}{2})$ and $G^{-}(l,m)=\sin^2(\frac{\pi (l+m)}{2})$. We are going to estimate the ratio $\gamma^+_l=D(l)_{Re}^{+}/D(l)_{Im}^{+}$ for $l+m=even$ components and $\gamma^-_l=D(l)_{Re}^{-}/D(l)_{Im}^{-}$ for $l+m=odd$ in order to investigate the statistical isotropy and Gaussianity of CMB signal. Effectively, we are thus only comparing power spectra composed of either symmetric ($+$) or asymmetric ($-$) terms, with respect to the S2 symmetry.

We have tested the $\gamma$-statistic for the WMAP 7 ILC data, until $l=100$, and compared them with 1000 Monte Carlo simulations. Below, in Fig.\ref{pach}, we plot the results of our analysis for both even and odd $l+m$ components of $C(l)$.
\FIGURE{
   \hbox{
    \includegraphics[scale=0.4]{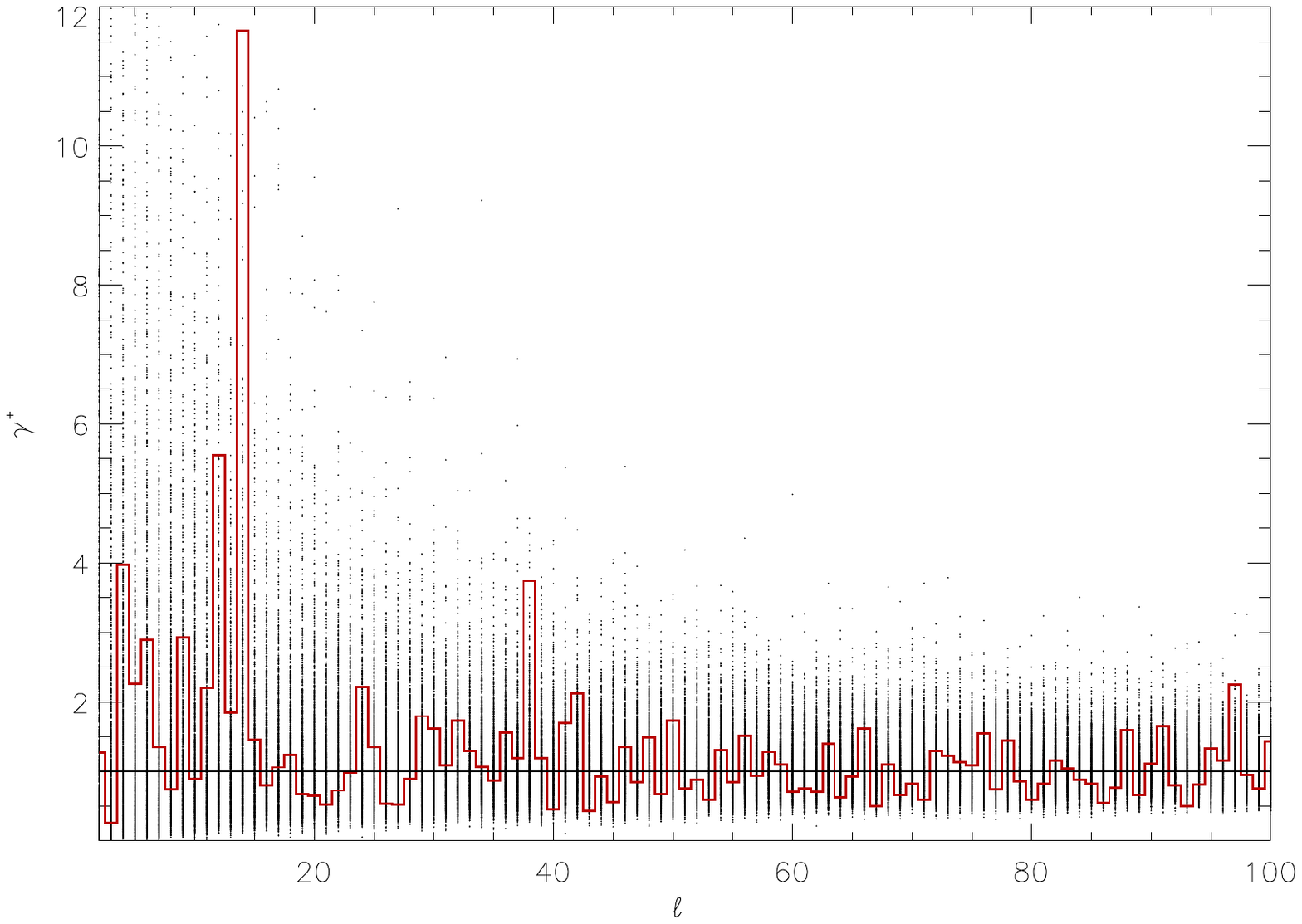}
\includegraphics[scale=0.4]{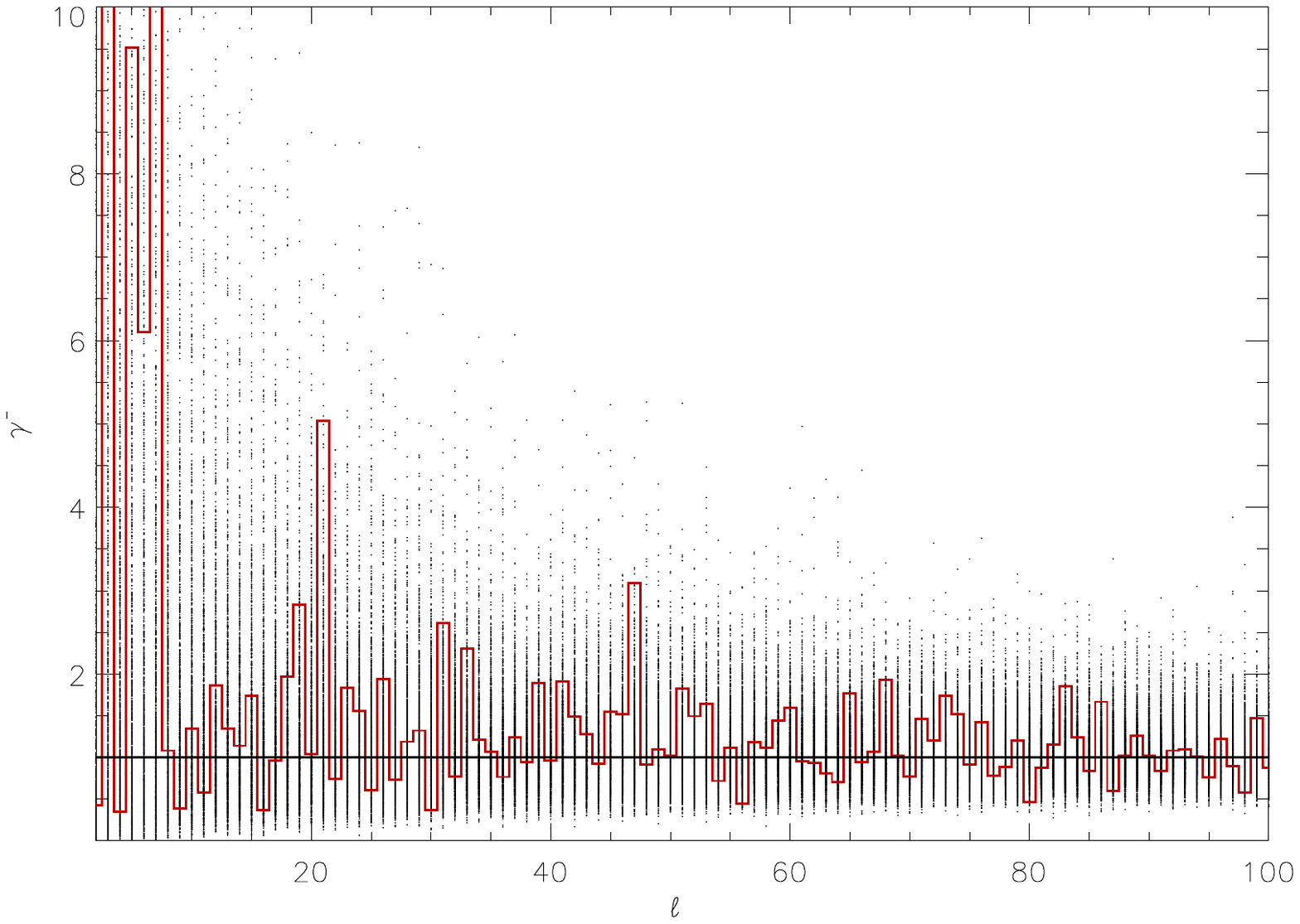}
     }
    \caption{Left panel. The ratio ($\gamma^{+}_{l}$) between $C(l)_{Re}$ and $C(l)_{Im}$ for even $l+m$: The red line shows the result from the ILC7 map, the black horizontal line shows the unit value, and the black dots correspond to the distribution of 1000 Monte Carlo simulations. Right panel. The same as left, but for $\gamma^{-}_{l}$.}
\label{pach}
}

\FIGURE{
   \hbox{
    \includegraphics[scale=0.17]{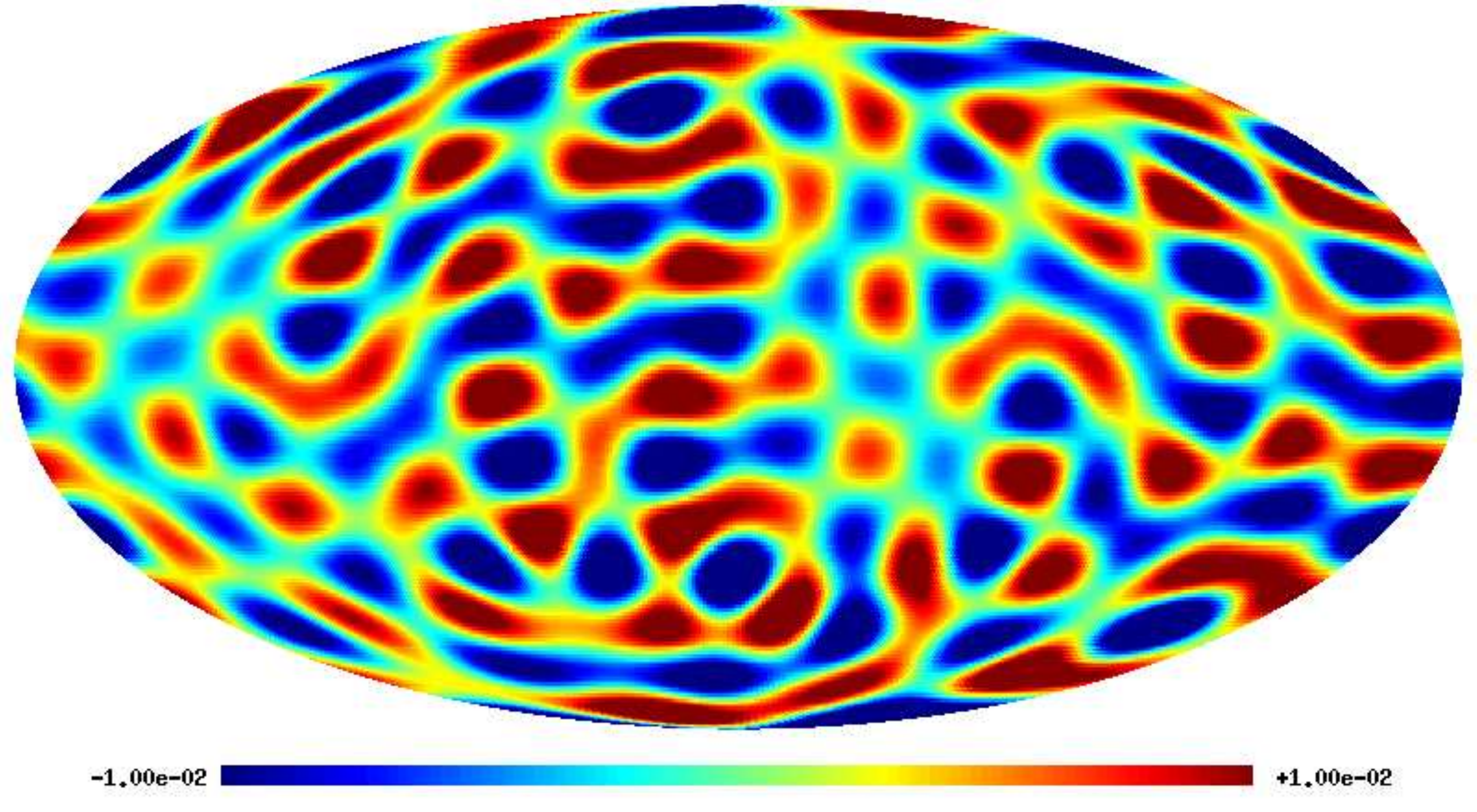}
    \includegraphics[scale=0.17]{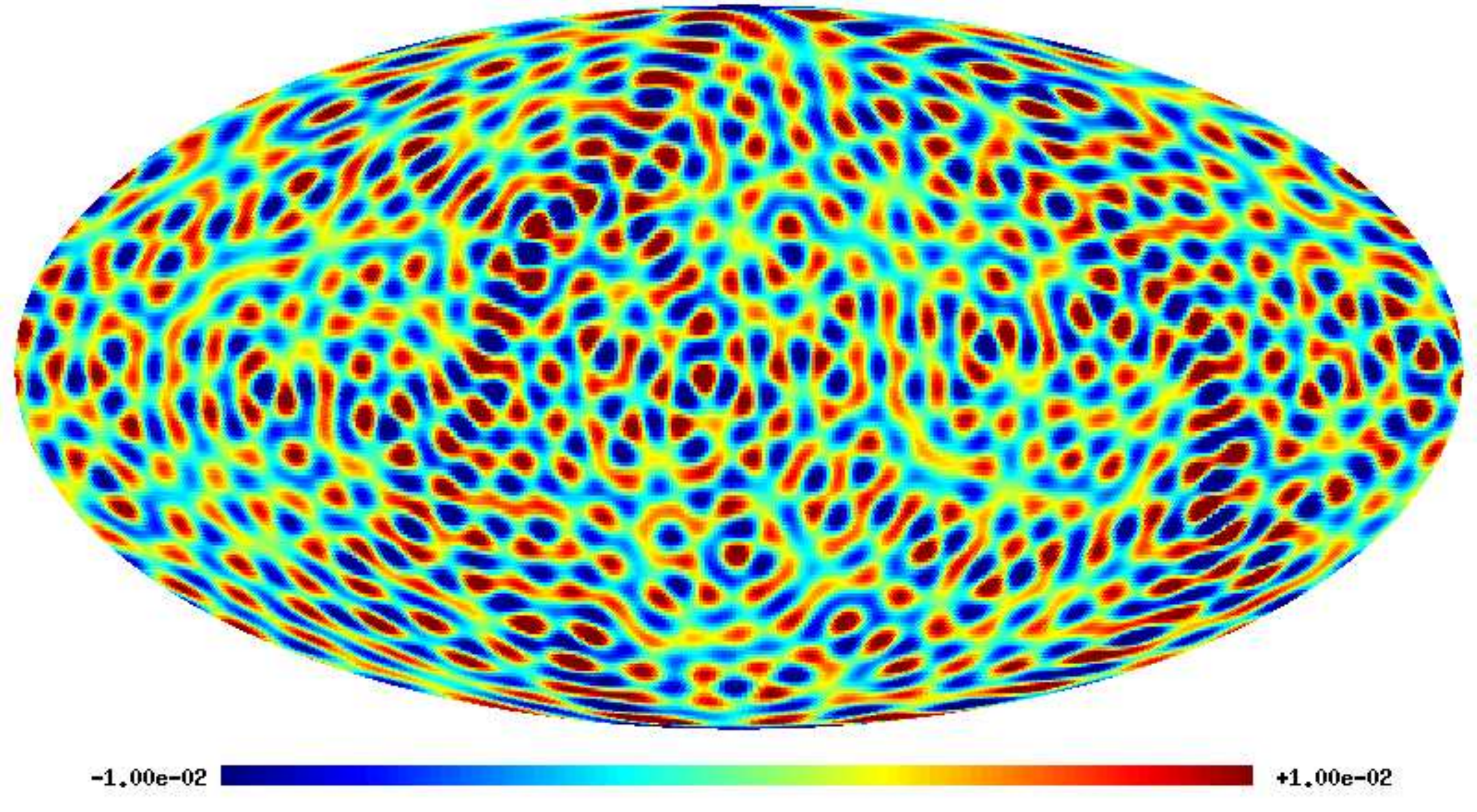}
    \includegraphics[scale=0.17]{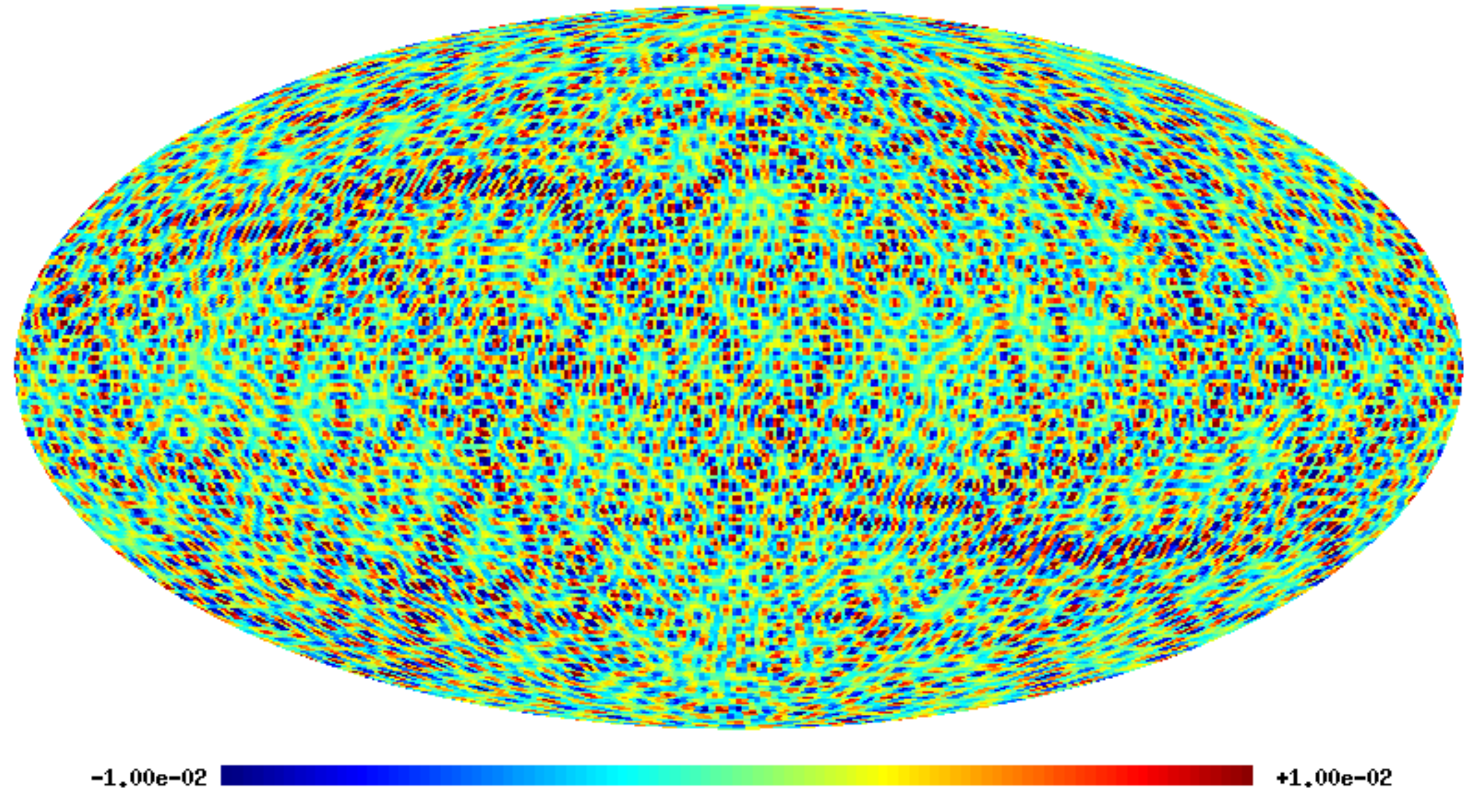}
     }
    \caption{The images of the WMAP ILC 7 components with $l=14$ (left), $l=38$ (middle), and $l=97$ (right panel)}
\label{f11}
}

\FIGURE{
   \hbox{
    \includegraphics[scale=0.17]{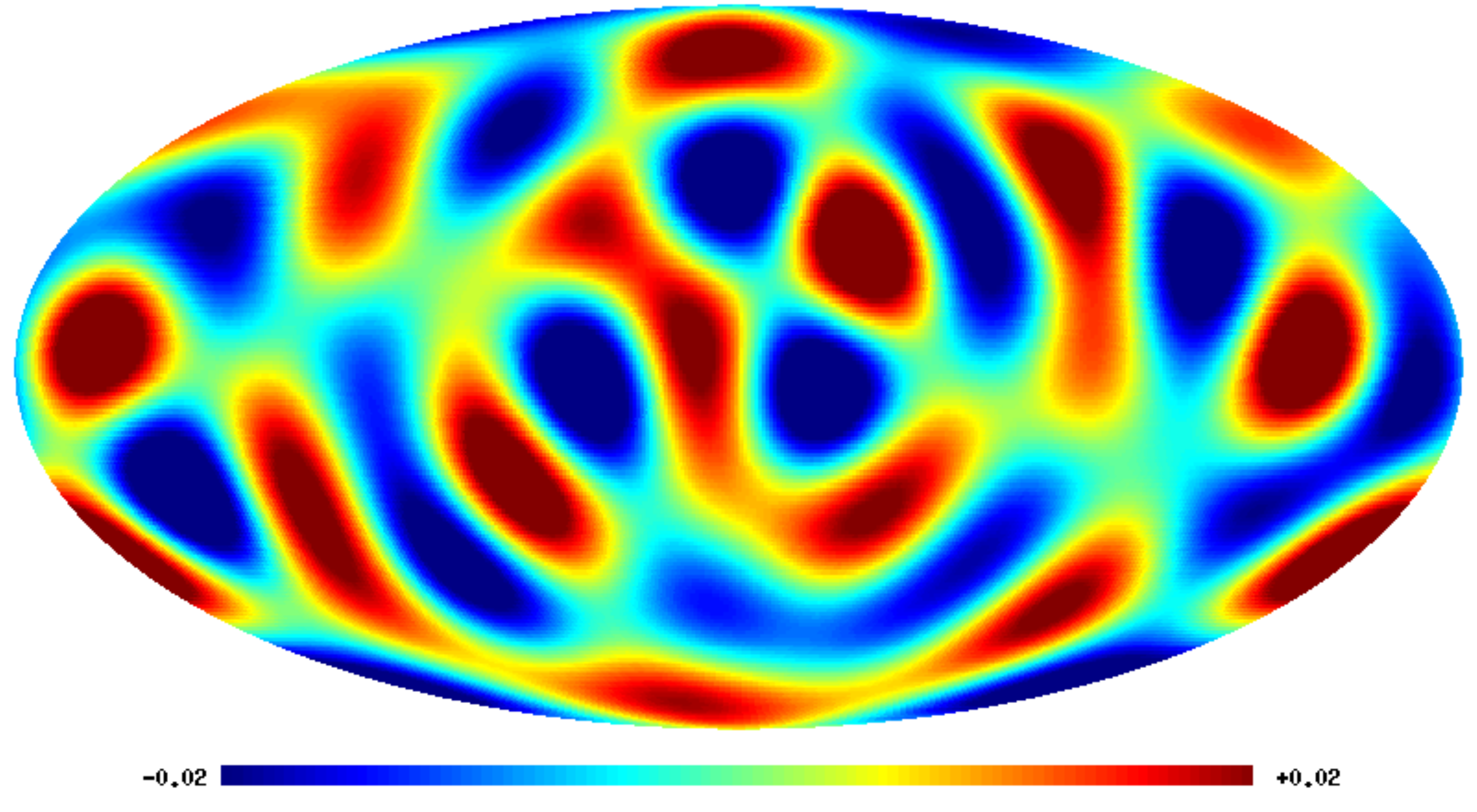}
    \includegraphics[scale=0.17]{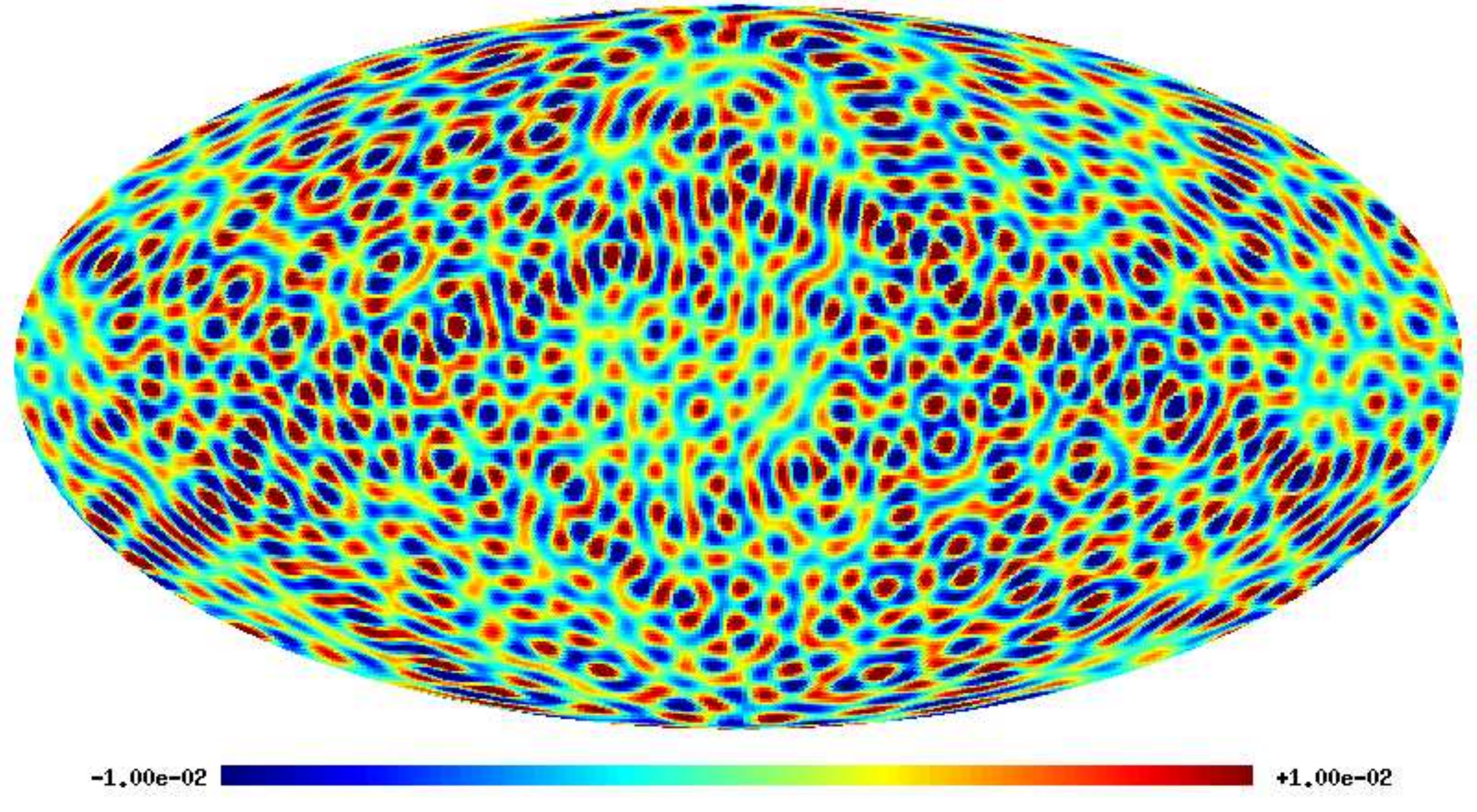}
    \includegraphics[scale=0.17]{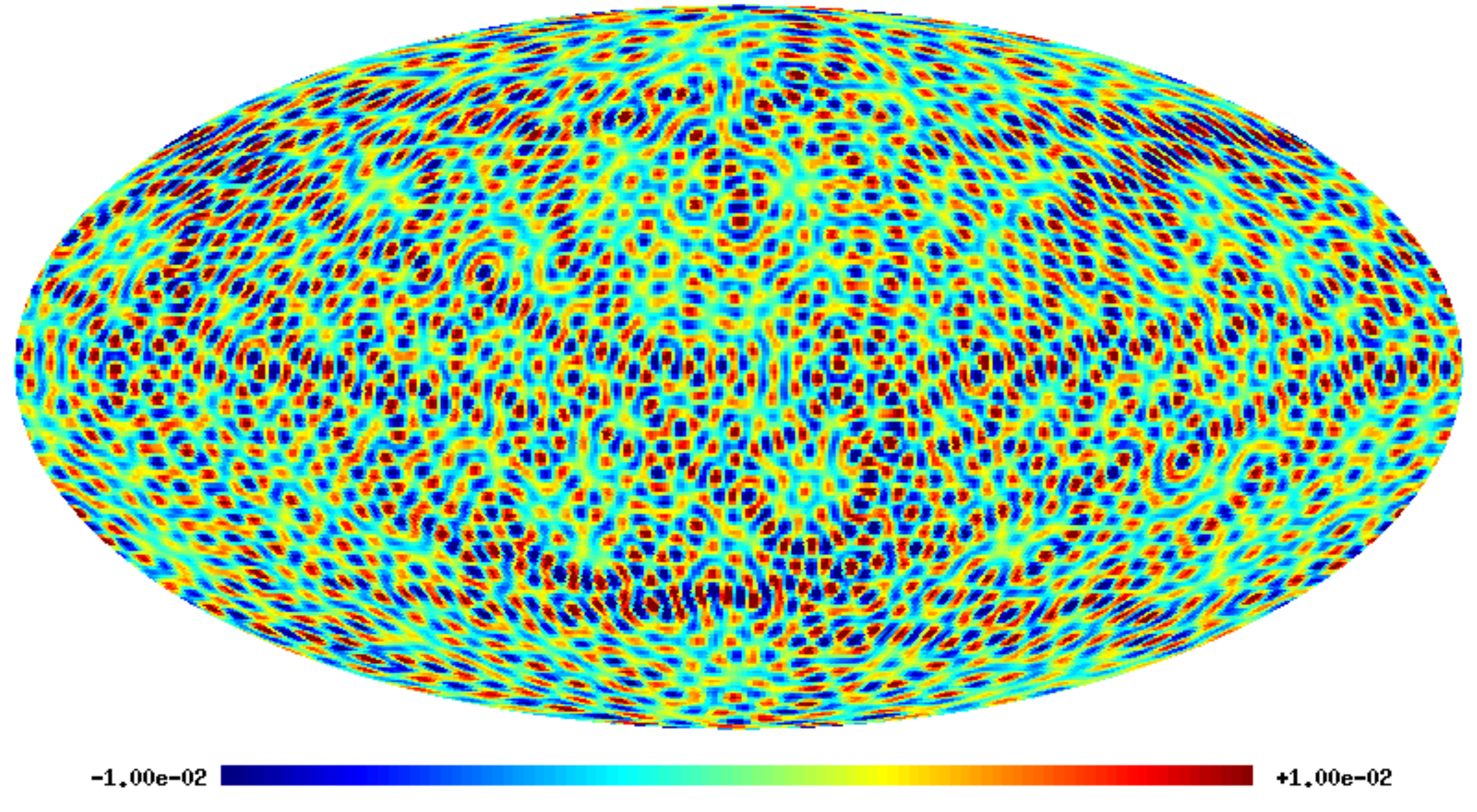}
     }
    \caption{The images of the WMAP ILC 7 components with $l=7$ (left), $l=47$ (middle), and $l=68$ (right panel)}
\label{f20}
}
As is evident, we have several peaks for the $\gamma^{+}_{l}$, which is most noticeable at $l=14$, where only 3 events out of 1000 realizations have values as high as the WMAP data. For that multipole the phase of $14,4$ component is $\psi_{14,4}=-3.061\simeq -\pi$, and $\psi_{14,10}= 2.920$, which indicate the most planar components. 
\FIGURE{
   \hbox{
    \includegraphics[scale=0.6]{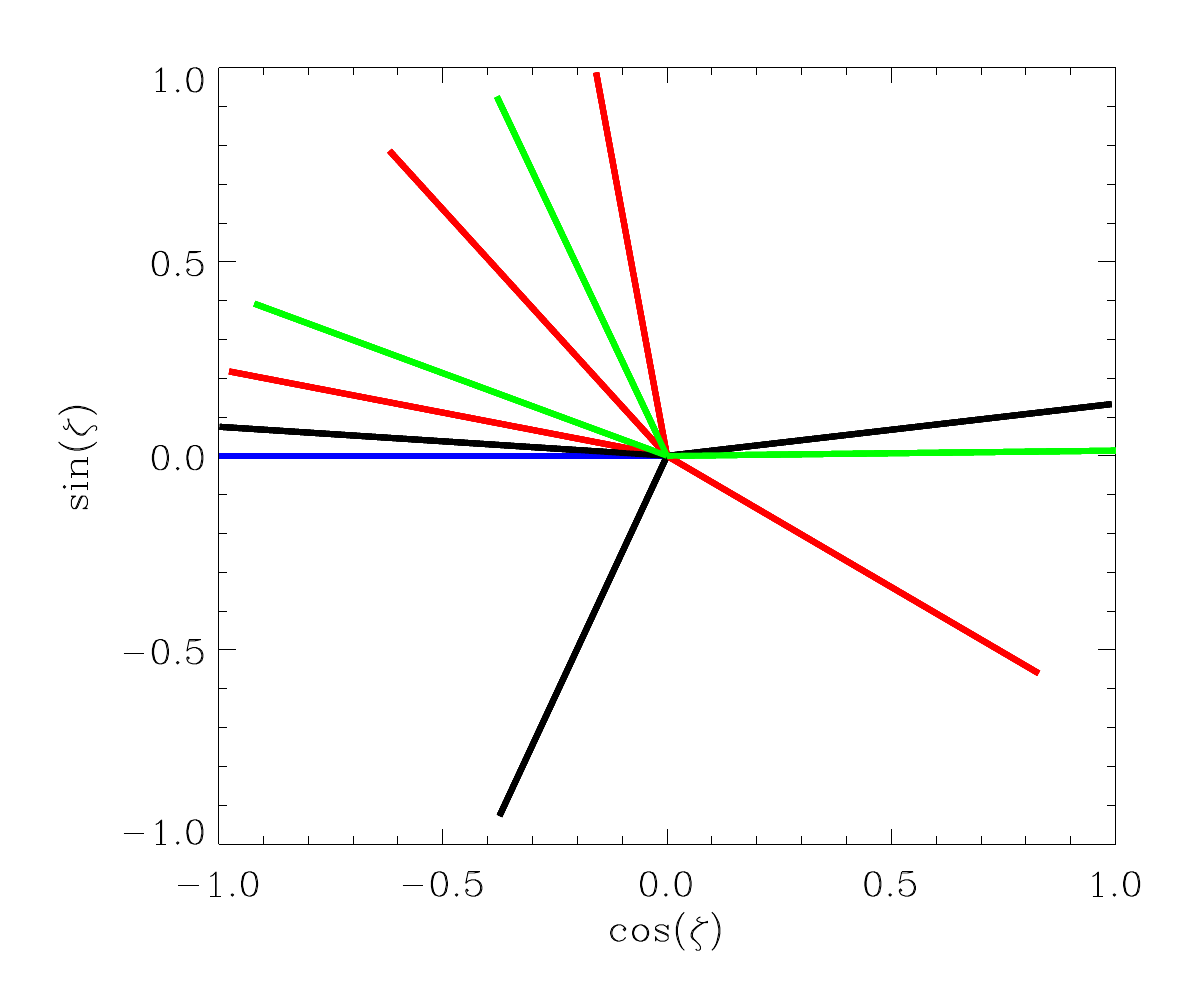}
    \includegraphics[scale=0.6]{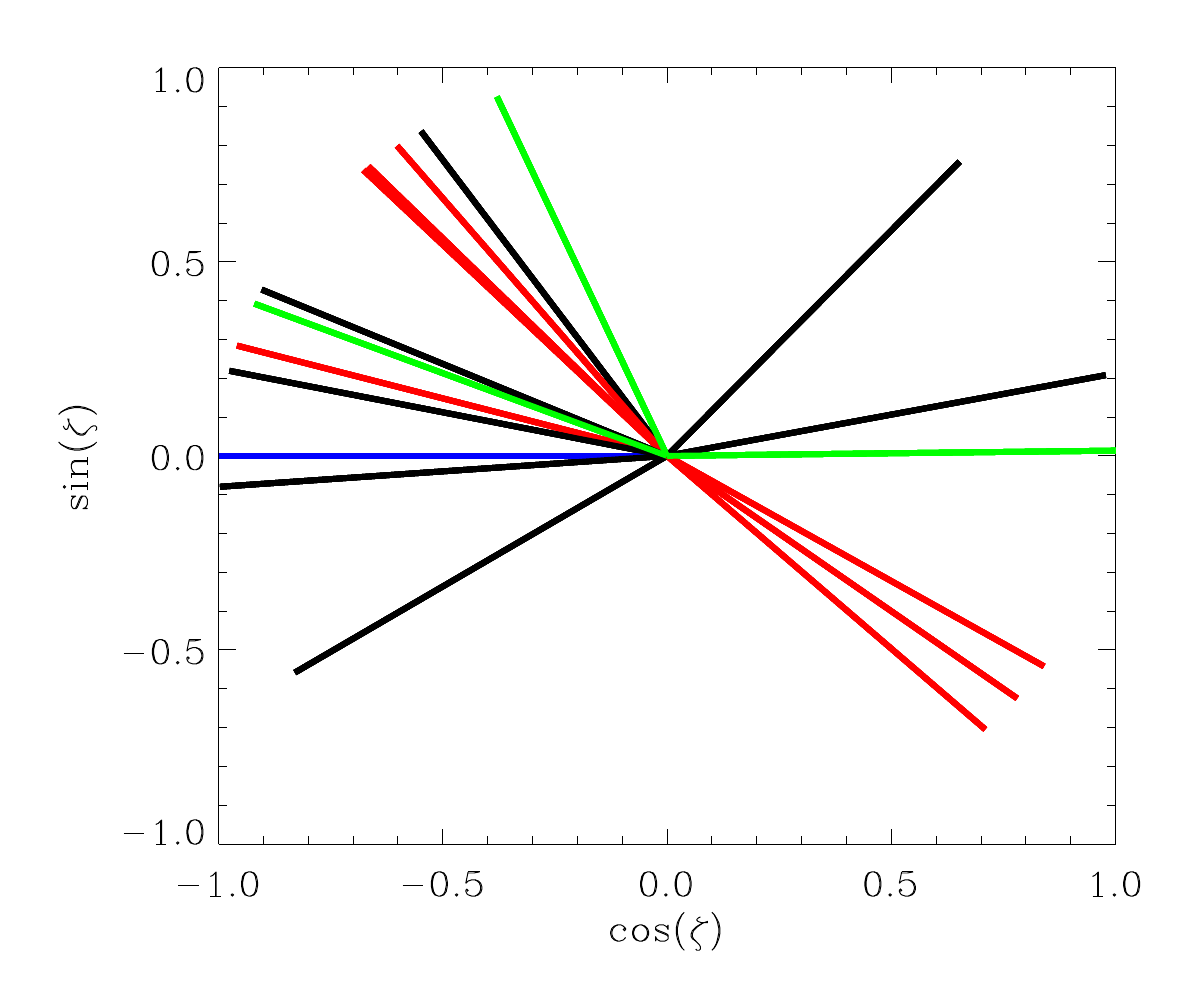}
     }
    \caption{The phases of the WMAP ILC 7 components with $l=7$ (left), and $l=14$ (right panel).
The blue line indicate $m=0$ component, the black lines correspond to $l+m=odd$, the red dash lines are for
$l+m=even$. The green lines show the phases of the octupole.}
\label{f22}
}

\TABLE{
\caption{ $\Re e$ and $\Im m$ parts (in $mK$) of the multipole coefficients $a_{7,m}$ for the ILC 7.}
\begin{tabular}{ccc}
\hline 
 $l=7,m$    &     $\Re e (a_{7,m})$ &$ \Im m(a_{7,m})$\\
\hline
$m=0$ & $ -5.159e-03 $& $0 $\\ 
$m=1$ & $ -1.409e-02 $ & $3.141e-03 $ \\ 
$m=2$ & $8.908e-03 $ & $1.199e-03$ \\ 
$m=3$ & $ -4.628e-03$ & $ 5.878e-03$ \\
$m=4$ & $-6.4957e-04 $ & $-1.6132e-03 $ \\
$m=5$ & $1.5803e-02 $ & $-1.069e-02 $ \\
$m=6$ & $-1.285e-02 $ & $9.648e-04 $ \\
$m=7$ & $-2.219e-03 $ & $1.387e-02 $ \\
\hline
\end{tabular}
\label{table2} }

 Other prominent peaks for $\gamma^{+}_{l}$ are found at $l=38$ and $l=97$, where only 6 and 3 events have value as large as WMAP data respectively. For the $\gamma^{-}_{l}$, we see a similar tendency: at $l=7$ we have 3 events higher than that of the WMAP data. Another notable peak is at $l=47$, where we have 4 events higher than the WMAP data.\\
The $\gamma^{-}_{l}$ have some very prominent peaks for $l < 10$ (at $l=3$ for instance, the ratio is in the order of $10^3$, and for $l=7$, the ratio is around $10^2$). These peaks are to high to include in the plot above, as they would have dwarfed the other values completely, and as we wanted the ratios to be easily comparable, we decided against a logarithmic y-axis. 
We summarize in Fig.\ref{f11}-\ref{f20} the images of the corresponding maps for the given values of $l$ and all $m$, for the most prominent peaks from Fig.\ref{pach}. One can clearly see the corresponding symmetries or asymmetries of these maps with respect to the galactic plane. For illustration of the morphology of these maps, in Fig.\ref{f22} we plot the phases of the $a_{l,m}$ for $l=7$ and $l=14$, combining $l+m=even$, and $l+m=odd$ modes. Even without special analysis, one can see that phases of $l+m=even$ modes are coherent with the phases of the octupole.

The peculiar behavior of the multipoles, detected by the $\gamma^+$ and $ \gamma^-$- tests, allow us to look closely at the corresponding symmetry estimators for each. In table \ref{table2} we show the corresponding real and imaginary parts of the  $a_{l,m}$-coefficients for $l=7$ peak. From this table, it is clearly seen that for $l=7$ the ratio $\alpha_{7,1}=\frac{\Im m(a_{7,5})}{\Im m(a_{7,6})}\simeq 11.08 $. This parameter is an analogue of the $\alpha_3$- parameter for the octupole and it corresponds to the $S3$-symmetry. Taking under consideration that the corresponding probability for that parameter is given by Eq \ref{prob} with $X=11.08$, we get $P(\alpha_{7,1})\simeq 0.028$. The analogue of the $\beta$-parameter of the octupole is $\beta_{7,4}=\frac{\Re e(a_{7,1})}{\Re e(a_{7,4})}\simeq 21.69$, with $P(\beta_{7,4})\simeq 0.015$. Thus, for $l=7$ the most established type of symmetry is S2, with an additional S3 component.

\section{Conclusion}
In this paper, we have introduced 3 symmetries on the sphere of the sky: one with respect to the antipodal points ($S1$), one with respect to the galactic plane ($S2$), and one where we rotate in the $\phi$-direction ($S3$). The $a_{l,m}$-coefficients, would either be symmetric or asymmetric with respect to the three symmetries, depending on the value of $l$, $l+m$ and the real and imaginary components of the $a_{l,m}$-components. Therefore we have introduced a symmetry test, based on the ratio between the symmetric and asymmetric $a_{l,m}$.

We have tested it for the WMAP7 octupole in particular, as it is the most powerful of the asymmetric multipoles, and the results were odd at the level of $0.58\%$ for $\alpha_{3}$, and $9.4\%$ for the $\alpha_{1}$ compared to 10000 Monte Carlo simulations. The quadru- and octupole have previously been extensively investigated in connection with the problem of alignment, but this result is only based on the octupole, unconnected with the quadropole. We have shown, that the octupole carry anomalously little power in the imaginary part of the most asymmetric component, $a_{3,2}$, especially compared with the imaginary part of the more symmetric $a_{3,3}$.\\
We have further introduced a symmetry test, based on the ratio between power spectra of real and imaginary parts of the $a_{l,m}$, created from sums over even and odd values of $l+m$ only ($\gamma^{+}_{l}$ and $\gamma^{-}_{l}$ respectively). We tested this for the WMAP7 data, in the range $2<l<100$, and compared with 1000 Monte Carlo simulation. We found notable deviations from the simulations at $l=14$ ($0.3\%$), $l=38$ ($0.6\%$) and $l=97$ ($0.3\%$) for $\gamma^{+}_{l}$, and at $l=7$ ($0.4\%$) and $l=47$ ($0.4\%$) for $\gamma^{-}_{l}$.

We have tested all mentioned anomalies by the phase analysis and find that all these symmetries and asymmetries are closely related to the correlations of the phases. We have investigated the coupling of the octupole phases with phases of dipole and foreground and confirm remarkable correlations of the octupole with the kinematic dipole. At the same time we note, that
residuals of the foregrounds could play a significant role in the formation of the peculiar symmetry of the octupole. We believe, that our symmetry test would be very useful for estimating the quality of the separation of the primordial CMB from non-cosmological signals, making the analysis of the primordial non-Gaussianity more sensitive.

\section{Acknowledgments}
We are grateful to the anonymous referee for very stimulating questions and remarks, and to Peter Coles for discussions. 
We acknowledge the use of the Legacy Archive for Microwave Background Data Analysis (LAMBDA). 
Our data analysis made use of the GLESP package \cite{Glesp}, and of HEALPix \cite{Healpix_primer}.
This work is supported in part by Danmarks Grundforskningsfond, which allowed the establishment of the Danish Discovery Center, and by FNU grant 272-06-0417, 272-07-0528 and 21-04-0355.

\end{document}